\pdfoutput=1
\documentclass[a4paper,5pt,intlimits,twopages]{iopart}
\expandafter\let\csname equation*\endcsname\relax
\expandafter\let\csname endequation*\endcsname\relax
\usepackage[tbtags]{amsmath}
\usepackage{amsthm,amssymb,dsfont,mathrsfs,bbm,xfrac,tabularx,booktabs}
\usepackage{cite}
\usepackage{mathtools}
\usepackage{multirow}
\usepackage{upgreek}
\usepackage[cal=esstix]{mathalfa}
\usepackage[usenames,dvipsnames]{xcolor}
\usepackage[caption=false]{subfig}
\usepackage[english]{babel}
\usepackage[export]{adjustbox}
\usepackage{geometry}
\usepackage{hyperref}
\hypersetup{pdfauthor={Piccioli Semerjian Sicuro Zdeborova},pdftitle={Aligning random graphs with a sub-tree similarity message-passing algorithm},%
            colorlinks, linktocpage=true, pdfstartpage=3, pdfstartview=FitV,%
    breaklinks=true, pdfpagemode=UseNone, pageanchor=true, pdfpagemode=UseOutlines,%
    plainpages=false, bookmarksnumbered, bookmarksopen=true, bookmarksopenlevel=1,%
    hypertexnames=true, pdfhighlight=/O,%
    urlcolor=orange, linkcolor=blue, citecolor=red, 
        }
\usepackage{tikz}
\usetikzlibrary{patterns,decorations,arrows,shapes.geometric,arrows,mindmap,backgrounds,positioning,positioning,fit,backgrounds,positioning,arrows,matrix,calc}
\usetikzlibrary{shapes,backgrounds,mindmap,shadows,shapes.geometric,topaths,snakes,arrows,pgfplots.groupplots,fadings,calc,decorations.markings,positioning,chains,fit}
\usepackage{pgfplotstable}

\usepackage{lmodern}
\usepackage[T1]{fontenc}

\newcommand{\bA}{{\bf A}}
\newcommand{\bB}{{\bf B}}
\newcommand{\bC}{{\bf C}}
\newcommand{\bG}{{\bf G}}

\newcommand{\bpis}{\boldsymbol{\pi_\star}}
\newcommand{\bpi}{\boldsymbol{\pi}}
\newcommand{\bpih}{\boldsymbol{\widehat{\pi}}}
\newcommand{\bPis}{\boldsymbol{\Pi_\star}}
\newcommand{\bPih}{\boldsymbol{\widehat{\Pi}}}
\newcommand{\pih}{\widehat{\pi}}
\newcommand{\Po}{{\rm Po} }
\newcommand{\ov}{{\rm ov}}
\newcommand{\lb}{{l_{\rm b}}}
\newcommand{\lr}{{l_{\rm r}}}
\newcommand{\lbi}{{l_{\rm bi}}}
\newcommand{\lbp}{{l'_{\rm b}}}
\newcommand{\lrp}{{l'_{\rm r}}}
\newcommand{\lbip}{{l'_{\rm bi}}}
\newcommand{\hq}{\widehat{q}}
\newcommand{\tq}{\widetilde{q}}
\newcommand{\dq}{\dot{q}}

\DeclareMathOperator{\dd}{{\mathrm d}}

\usepackage{tikz}
\usepackage{pgfplots}
\usepgfplotslibrary{fillbetween}
\usetikzlibrary{calc,shapes.multipart,patterns}
\tikzset{
bicolor/.style 2 args={
  dashed,dash pattern=on 2pt off 2pt,#1,thick,
  postaction={draw,dashed,dash pattern=on 2pt off 2pt,#2,dash phase=2pt,thick}
  },
}

\newcommand{\comprimi}{\medmuskip=0mu
\thinmuskip=0mu
\thickmuskip=0mu}

\allowdisplaybreaks

\usepackage{etoolbox}
\makeatletter
\newrobustcmd{\fixappendix}{%
  \patchcmd{\l@section}{1.5em}{7em}{}{}%
  \patchcmd{\l@subsection}{2.3em}{7em}{}{}%
}
\makeatother

\begin{document}
\title{Aligning random graphs with a sub-tree similarity message-passing algorithm}

\author{Giovanni Piccioli$^{1}$, Guilhem Semerjian$^2$, Gabriele Sicuro$^3$, Lenka Zdeborov\'a$^{1}$}
\address{$^1$SPOC Lab, EPFL, Rte Cantonale, 1015 Lausanne, Switzerland}
\address{$^2$Laboratoire de Physique de l'\'Ecole Normale Sup\'erieure, ENS, Universit\'e PSL, CNRS, Sorbonne Universit\'e, Universit\'e Paris Cit\'e, F-75005 Paris, France}
\address{$^3$ King's College London, Strand, WC2R 2LS London, United Kingdom}
\begin{abstract}
The problem of aligning Erd\H os--R\'enyi random graphs is a noisy, average-case version of the graph isomorphism problem, in which a pair of correlated random graphs is observed through a random permutation of their vertices. We study a polynomial time message-passing algorithm devised to solve the inference problem of partially recovering  the hidden permutation, in the sparse regime with constant average degrees. We perform extensive numerical simulations to determine the range of parameters in which this algorithm achieves partial recovery. We also introduce a generalized ensemble of correlated random graphs with prescribed degree distributions, and extend the algorithm to this case.
\end{abstract}


\section{Introduction}
The \textit{graph alignment problem} (GAP) is a classical combinatorial optimization problem consisting in finding a bijection between the vertex sets of two graphs in such a way that their edge sets are maximally aligned. To make this statement more precise, let us denote $A\in\mathds R^{n\times n}$ and $B\in\mathds R^{n\times n}$ the (possibly weighted) adjacency matrices of two graphs having the same number $n$ of vertices. The goal is to find the permutation $\hat \pi\in\mathcal S_n$, with $\mathcal S_n$ the set of the permutations of $n$ elements, such that $\hat \pi=\arg\max_{\pi\in\mathcal S_n}\sum_{i<j}A_{ij}B_{\pi(i)\pi(j)}$. The wide interest in this problem is due to the large number of applications involving the solution of a GAP, from pattern recognition~\cite{Conte2004} to network de-anonymization~\cite{Narayanan2008,Pedarsani2011} or alignment of molecular and protein-interaction networks in biology \cite{ramani2003exploiting,berg2006cross,li2007alignment,singh2008global}. The GAP has also been used as a prototypical and challenging problem to evaluate the performance of graph neural networks \cite{nowak2018revised,azizian2020expressive}.

In the general formulation above the GAP is also known under the name of \textit{quadratic assignment problem}~\cite{Burkard1998}, which belongs to the computational class of NP-hard problems (although some special settings allow for a polynomial-time solution~\cite{Burkard1998}). This worst-case hardness result leaves open the possibility that some ``typical'' instances are efficiently solvable. To give a precise meaning to this notion of typicality a number of  studies focused therefore on the alignment of pairs of graphs obtained from some random ensembles. In these ensembles, each pair is generated with the same vertex set of cardinality $n$ according to probabilistic rules implying some correlations between the two graphs, then the information of the vertex correspondence is removed by a random reshuffling $\bpis$ of the labels of one of the graphs. As a result, this \textit{planted} GAP takes the form of an inference problem in which the planted permutation $\bpis$ has to be, at least approximately, recovered. The goal of recovering the permutation $\bpis$ rather than maximally aligning the graphs also stems from applications where such a ground truth permutation is often assumed to exist, and the inference (rather than the optimization) version of the GAP problem is thus our interest in this paper.  We may wonder at this point if the exact, or perfect, recovery of $\bpis$ is achievable with high probability over the ensemble samples. We can also ask, less ambitiously, if a \textit{partial} recovery of $\bpis$ is feasible with finite probability, i.e., if it is possible to recover the correct matching of a finite fraction of vertices. Another question concerns the possibility of detecting the correlations between the graphs, namely to distinguish between a sample of the correlated ensemble and one made of two independent graphs. These questions have been studied in a series of theoretical works, for different relevant graph ensembles, in the limit of large graph sizes $n\to +\infty$.

In this contribution we will mostly focus on the correlated Erd\H os--R\'enyi ensemble $\mathrm{G}(n,\sfrac{\lambda}{n},s)$. This ensemble, that we will detail in Section~\ref{sec:er}, was introduced in~\cite{Pedarsani2011} in the context of de-anonymization of social networks. Here we anticipate that an element of this ensemble is given by a pair of correlated Erd\H os--R\'enyi graphs, both with the same average degree $\lambda$, on the same set of $n$ vertices: the parameter $0\leq s\leq 1$ measures the degree of correlation, so that $s=1$ corresponds to the case of identical graphs and $s=\lambda / n$ corresponds to a pair of independently generated Erd\H os--R\'enyi graphs. After the generation, the labels of one of the two graphs are reshuffled by a permutation $\bpis$ to be recovered. The answers to the questions raised above on the possibility of exact recovery, partial recovery, and detection, depend on the scaling of the parameters $\lambda$ and $s$ with the size $n$ of the graphs. Cullina and Kiyavash~\cite{Cullina2016} studied the exact recovery question in the regime of diverging degrees, showing that it is possible to exactly recover $\bpis$ if and only if $\lambda s-\ln n\to+\infty$ as ${n\to+\infty}$. The range of parameters for which polynomial time algorithms succeed in this exact recovery task have been progressively improved in~\cite{Ding2018,Fan2019,MaRuTi21}, covering the case where the average degree $\lambda$ is slightly greater than $\ln n$ and $s$ is a constant sufficiently close to $1$. On the other hand, it is impossible to exactly recover $\bpis$ for $\lambda=O(1)$~\cite{Cullina2016}. For this reason the authors of~\cite{GaMa20,WuXuYu21,GaMaLe21a,GaMaLe21b} focused in this regime on the possibility of a \textit{partial} recovery of the true labelling. In~\cite{GaMaLe21a} it is shown that the fraction of correctly matched pairs of vertices is upper-bounded by $c(\lambda s)$, the largest non-negative solution of the equation $1-c=\e^{-\lambda s c}$, for any statistical estimator. This implies that for $\lambda s\leq 1$ partial recovery is \textit{information-theoretically infeasible}, i.e., for this set of parameters it is not possible to recover $\bpis$, not even partially. On the other hand, in~\cite{WuXuYu21} it is proven instead that for $\lambda s>4$ partial recovery is information-theoretically feasible (improving on a previous bound in~\cite{Hall2020}). It has also been shown in~\cite{GaMa20,GaMaLe21b} that there exists a \textit{polynomial-time feasible phase} in the region $\lambda s>1$ (for large enough values of $s$). All these results leave open a complete determination of the phase diagram of the partial recovery problem in the $(\lambda,s)$ plane, namely the boundaries of the impossible, easy (meaning feasible in polynomial time) and hard (information-theoretically feasible but in an a priori exponential time under some computational complexity hypothesis) phases. The existence of an hard phase for large enough $\lambda$ follows from the bounds of \cite{WuXuYu21,GaMaLe21b}, but its precise boundary is not accurately known.

Let us also mention other ensembles on which the graph alignment problem has been studied. Correlated random graphs with a hidden community structure have been considered in~\cite{OnGaEr16,RaSr21} in the context of network de-anonymization problems. They are built by first generating a ``parent graph'' using the stochastic block model (SBM), in which the vertices bear some labels interpreted as communities, the probability of presence of an edge between two vertices depending on their labels. Then two graphs are obtained from it by randomly removing edges from the parent one with probability $1-s$, independently for each of the two copies, and reshuffling the vertex labels of one of the graphs. The information-theoretical possibility of exact recovery in this ensemble has been characterized in~\cite{RaSr21} for the regime of degrees logarithmic in $n$. Another variant of the problem concerns the alignment of correlated random matrices, which corresponds to the case of weighted complete graphs~\cite{Ding2018,GaLeMa19,Fan2019,WuXuYu21,Ga20}: one draws a pair of correlated random matrices $(\bA,\bC)$ such that, independently for all $i<j$, $\bA_{ij}$ and $\bC_{ij}$ are standard Gaussian variables with correlation coefficient $s\in[0,1]$. The observed pair is obtained by reshuffling the rows and columns of one of the matrices, $\bC_{ij}=\bB_{\bpis(i)\,\bpis(j)}$, and the goal is to recover the uniformly random permutation $\bpis$ from the observation of $(\bA,\bB)$. In~\cite{Ding2018,Fan2019} it is shown that a spectral (polynomial-time) algorithm exactly recovers $\bpis$ if $\sqrt{1-s^2}=O(\ln^{-1}n)$. On the other hand, in~\cite{Ga20,WuXuYu21} it is proven that it is information-theoretically possible to recover $\bpis$ if $s^2\geq \frac{c\ln n}{n}$ for some $c>4$: this suggests that in this problem there might be a wide \textit{hard phase} in which polynomial-time algorithms cannot exactly recover $\bpis$.

In this paper we develop a message-passing strategy for the graph alignment problem on correlated random graphs of constant degrees. It is substantially different from the message-passing algorithms of~\cite{BrBrMaTrWeZe10,BaGlSaWa13}, that were based on a belief-propagation approximation of the posterior distribution on the unknown permutation, and whose convergence required some side-information or the use of sophisticated numerical tricks (decimation, reinforcement and/or the introduction of biases). Instead we use crucially the locally tree-like character of the random graphs to compute a \textit{score} for each pair of vertices of the two graphs to align based on the similarity of their neighborhoods, that quantifies how likely two vertices were matched through $\bpis$ and allows thus to build an estimator of this unknown permutation achieving partial recovery in some portion of the parameter space. While we were working on this project we became aware of the independent work of~\cite{GaMaLe21b}, which followed a very similar reasoning and contains an essentially equivalent algorithm. Our derivation is, however, slightly different, and we believe the extensive numerical simulations we present are a useful complement to the rigorous bounds of~\cite{GaMaLe21b}.

The rest of the paper is organized as follows. In Section~\ref{sec:corrgraphs} we introduce the ensemble $\mathrm{G}(n,\sfrac{\lambda}{n},s)$ that is going to be the main object of our analysis, and we discuss the local properties of pairs of correlated graphs drawn from this ensemble. In Section~\ref{sec:inference} we present the Bayesian formulation of the inference problem and the message-passing algorithm obtained via a truncation of the posterior distribution. The results of the numerical experiments on this algorithm are discussed in Section~\ref{sec:numerical}, along with comparisons with the known theoretical bounds. Section~\ref{sec:tree} is devoted to the liming tree problem that arises from the analysis of the algorithm. Finally, in Section~\ref{sec:conclusions} we draw our conclusions. In \ref{app:genensembles} we introduce a generalized ensemble of correlated random graphs with prescribed degree distributions and extend the message-passing algorithm to this ensemble (considering also the case of weighted graphs). Further results of the numerical experiments are presented in \ref{app:numerical}.

Throughout the paper we will denote $[l]\coloneqq \{1,\dots,l\}$, and we will use bold fonts for random variables, a notation that we anticipated throughout this introduction.

\section{Correlated Erd\H os-R\'enyi random graphs}
\label{sec:corrgraphs}

\subsection{Definition}
\label{sec:er}

Let us start by defining the correlated Erd\H os-R\'enyi (ER) random graph ensemble~\cite{Pedarsani2011} denoted $\mathrm{G}(n,\sfrac{\lambda}{n},s)$, that depends on two real parameters, $\lambda > 0$ and $s\in[0,1]$. An element of this ensemble is a pair $(\bA,\bC)$ of random graphs on a common vertex set $V = [n]$, defined by their adjacency matrices $\bA$ and $\bC$ (here and in the following we use for simplicity the same notation for a graph and its adjacency matrix) generated as follows: independently for each of the $\frac{n}{2}(n-1)$ pairs of vertices $i<j$,
\begin{itemize}
\item $\bA_{ij}=\bC_{ij}=1$ with probability $\frac{\lambda}{n}s$;
\item $\bA_{ij}=1$, $\bC_{ij}=0$ with probability $\frac{\lambda}{n}(1-s)$;
\item $\bA_{ij}=0$, $\bC_{ij}=1$ with probability $\frac{\lambda}{n}(1-s)$;
\item $\bA_{ij}=\bC_{ij}=0$ with probability $1-\frac{\lambda}{n}(2-s)$.
\end{itemize}
The diagonal elements of the adjacency matrices are set to zero, $\bA_{ii}=\bC_{ii}=0$ (there is no self-loop in the graphs), and the adjacency matrices are completed by symmetry, $\bA_{ji}=\bA_{ij}$ and $\bC_{ji}=\bC_{ij}$ (the graphs are undirected). 

The marginal distributions of $\bA$ and $\bC$ are easily seen to coincide with the usual ER ensemble $\mathrm G(n,p=\sfrac{\lambda}{n})$, where each possible edge is present with probability $p$. The parameter $s\in[0,1]$ controls the correlation between the graphs $\bA$ and $\bC$: for $s=1$, they are strictly identical, $\bA=\bC$, whereas for $s=\frac{\lambda}{n}$ they are independent. As we will concentrate on the large size limit $n \to \infty$ with $\lambda$ and $s$ kept fixed any $s>0$ thus corresponds to a (positively) correlated situation.

We finally introduce a random permutation $\bpis$ uniformly drawn from the symmetric group $\mathcal S_n$, and define a graph $\bB$ as the image of $\bC$ through the reshuffling of its vertices' labels by the permutation $\bpis$. More explicitly, we define the adjacency matrix $\bB$ as $\bB_{ij}=\bC_{\bpis^{-1}(i),\bpis^{-1}(j)}$, or equivalently $\bC_{ij}=\bB_{\bpis(i),\bpis(j)}$. The marginal distribution of $\bB$ is then of course the same as $\bA$ and $\bC$, namely $\mathrm G(n,p=\sfrac{\lambda}{n})$.

\subsection{Local properties}
\label{sec:local_props}

\subsubsection{Single ER graphs}

The analysis of the graph alignment problem presented in the rest of the paper will rely crucially on the local properties of the correlated random graph ensemble that we shall now present. As a first step let us discuss the case of a single ER graph drawn from the $\mathrm G(n,p=\sfrac{\lambda}{n})$ ensemble (i.e., considering only $\bA$, $\bB$ or $\bC$). It is well-known that these graphs locally converge, in the large size limit $n\to\infty$, to Galton-Watson (GW) random trees with an offspring distribution given by the Poisson law of parameter $\lambda$, that we shall denote in the following $\Po(\lambda;l) = \e^{-\lambda} \frac{\lambda^l}{l!}$. To spell out more precisely the meaning of this statement we shall denote $\bA_{i,d}$ the subgraph of $\bA$ obtained by retaining the vertices that are at a distance smaller or equal than $d$ from the reference vertex $i$, where the distance between two vertices is the minimal number of edges on a path linking them. For an arbitrary choice of $i$, and for any fixed $d$, $\bA_{i,d}$ is a tree $T$ rooted in $i$ with high probability (w.h.p.), meaning with a probability going to 1 when $n \to \infty$. Moreover the law of $T$ corresponds to the first $d$ generations of a GW branching process: the root $i$ has a number $l$ of descendents drawn from the probability law $\Po(\lambda;l)$, each of them having an independent number of offsprings with the same law, and so on and so forth until the $d$-th generation has been reached (or until the branching process gets extinct). We shall denote $\mathbb{P}_0^{(d)}[T]$ the probability that a given tree $T$ is generated in this way, that admits the following recursive decomposition:
\begin{equation}
    \mathbb{P}_0^{(d)}[T] = \Po(\lambda;l) \prod_{j=1}^l \mathbb{P}_0^{(d-1)}[T_j] \ ,
\label{eq_P0}
\end{equation}
for a tree $T$ whose root has $l$ descendents which are themselves the roots of the $l$ subtrees $(T_1,\dots,T_l)$, and with the convention $\mathbb{P}_0^{(0)}=1$. To be precise, here and in the following all trees are understood to be rooted and labelled, and we consider two trees as equal if they are related by a relabelling that preserves the parent-offspring relationships.

Without entering into a formal proof of this local convergence property let us sketch its justification, which will be useful to address the generalization to correlated pairs of graphs. As the edges of an ER graph are independently present with probability $\lambda/n$, the degree of a vertex $i$ has a binomial distribution with parameters $(n-1,\lambda/n)$, that converges to $\Po(\lambda)$ as $n \to \infty$. Once the edges absent and present around $i$ have been revealed, one can continue the exploration process by exposing the edges adjacent to the neighbors of $i$, let us call them $\{i_1,\dots,i_l\}$. The number of neighbors of $i_1$ distinct from $i$ is binomial with parameters $(n-2,\lambda/n)$, that again converges to $\Po(\lambda)$ as $n \to \infty$. Moreover the probability that there is an edge between any two vertices in $\{i_1,\dots,i_l\}$ is of order $1/n$, as there is a finite number of possible edges between them, each being present with probability $\lambda/n$. This reasoning can be extended to the exploration of the neighborhood of $i$ up to any fixed distance $d$, as the number of revealed vertices remains finite while $n\to\infty$, hence all binomial of parameters $(n-o(n),\lambda/n)$ converge to $\Po(\lambda)$, and the probability of an edge being present among a fixed number of vertices being of order $1/n$.

\subsubsection{Aligned pairs of graphs}
\label{sec_aligned}

\begin{figure}
    \centering
    \begin{tikzpicture}[scale=0.75]

\draw[thick,blue,dash pattern= on 3pt off 5pt] (0,0) -- (-3,-2);
\draw[thick,red,dash pattern= on 3pt off 5pt,dash phase=4pt] (0,0) -- (-3,-2);
\draw[thick,blue,dash pattern= on 3pt off 5pt] (0,0) -- (-1,-2);
\draw[thick,red,dash pattern= on 3pt off 5pt,dash phase=4pt] (0,0) -- (-1,-2);
\draw[thick,blue] (0,0) -- (0,-2);
\draw[thick,blue] (0,0) -- (1,-2);
\draw[thick,red] (0,0) -- (2,-2);

\draw[thick,blue,dash pattern= on 3pt off 5pt] (-3,-2) -- (-4,-4);
\draw[thick,red,dash pattern= on 3pt off 5pt,dash phase=4pt] (-3,-2) -- (-4,-4);
\draw[thick,red] (-3,-2) -- (-3.25,-4);
\draw[thick,red] (-3,-2) -- (-2.5,-4);

\draw[thick,blue] (-1,-2) -- (-1.5,-4);
\draw[thick,red] (-1,-2) -- (-0.5,-4);

\draw[thick,blue,dash pattern= on 3pt off 5pt] (1,-2) -- (.5,-4);
\draw[thick,red,dash pattern= on 3pt off 5pt,dash phase=4pt] (1,-2) -- (.5,-4);
\draw[thick,red] (1,-2) -- (1.5,-4);

\draw[thick,red] (2,-2) -- (2.5,-4);

\fill[black] (0,0) circle (2pt) node [above] {1};
\fill[black] (-3,-2) circle (2pt) node [above left] {3};
\fill[black] (-1,-2) circle (2pt) node [above left] {6};
\fill[black] (0,-2) circle (2pt) node [below] {7};
\fill[black] (1,-2) circle (2pt) node [above right] {9};
\fill[black] (2,-2) circle (2pt) node [above right] {5};
\fill[black] (-4,-4) circle (2pt) node [below] {4};
\fill[black] (-3.25,-4) circle (2pt) node [below] {18};
\fill[black] (-2.5,-4) circle (2pt) node [below] {11};
\fill[black] (-1.5,-4) circle (2pt) node [below] {13};
\fill[black] (-.5,-4) circle (2pt) node [below] {15};
\fill[black] (1.5,-4) circle (2pt) node [below] {16};
\fill[black] (0.5,-4) circle (2pt) node [below] {19};
\fill[black] (2.5,-4) circle (2pt) node [below] {12};

\draw[-latex] (3.5,-2) -- (4.5,-2);

\begin{scope}[xshift=8cm]

\draw[thick,blue] (0,0) -- (-2,-2);
\draw[thick,blue] (0,0) -- (-1,-2);
\draw[thick,blue] (0,0) -- (0,-2);
\draw[thick,blue] (0,0) -- (1,-2);

\draw[thick,blue] (-2,-2) -- (-2,-4);

\draw[thick,blue] (-1,-2) -- (-1,-4);

\draw[thick,blue] (1,-2) -- (1,-4);

\fill[black] (0,0) circle (2pt) node [above] {1};
\fill[black] (-2,-2) circle (2pt) node [above left] {3};
\fill[black] (-1,-2) circle (2pt) node [above left] {6};
\fill[black] (0,-2) circle (2pt) node [below] {7};
\fill[black] (1,-2) circle (2pt) node [above right] {9};
\fill[black] (-2,-4) circle (2pt) node [below] {4};
\fill[black] (-1,-4) circle (2pt)node [below] {13};
\fill[black] (1,-4) circle (2pt) node [below] {19};

\end{scope}

\begin{scope}[xshift=13cm]

\draw[thick,red] (0,0) -- (-1,-2);
\draw[thick,red] (0,0) -- (0,-2);
\draw[thick,red] (0,0) -- (1,-2);

\draw[thick,red] (-1,-2) -- (-2.5,-4);
\draw[thick,red] (-1,-2) -- (-1.75,-4);
\draw[thick,red] (-1,-2) -- (-1,-4);

\draw[thick,red] (0,-2) -- (0,-4);

\draw[thick,red] (1,-2) -- (1,-4);

\fill[black] (0,0) circle (2pt) node [above] {1};
\fill[black] (-1,-2) circle (2pt) node [above left] {3};
\fill[black] (0,-2) circle (2pt) node [above left] {6};
\fill[black] (1,-2) circle (2pt) node [above right] {5};
\fill[black] (-2.5,-4) circle (2pt) node [below] {4};
\fill[black] (-1.75,-4) circle (2pt) node [below] {18};
\fill[black] (-1,-4) circle (2pt) node [below] {11};
\fill[black] (0,-4) circle (2pt) node [below] {15};
\fill[black] (1,-4) circle (2pt) node [below] {12};

\end{scope}

\end{tikzpicture}

    \caption{Left: an example of the colored GW tree $\mathcal T$ obtained by a local exploration of the aligned graphs represented by $\bG$, starting from the vertex $i=1$, up to distance $2$. Right: the corresponding neighborhoods of $i=1$ in the two graphs $\bA$ and $\bC$, $(T,T')=(b({\mathcal T}),r({\mathcal T}))$. Note that the red and bicolored edges below the vertex $9$ do not appear in the red tree, because this vertex has been reached by a blue edge in $\mathcal T$. The neighborhoods of $i$ in $\bA$ and $i'=\bpis(i)$ in $\bB$ are isomorphic to these two trees, with a random relabelling of the vertices.}
    \label{fig:tree}
\end{figure}
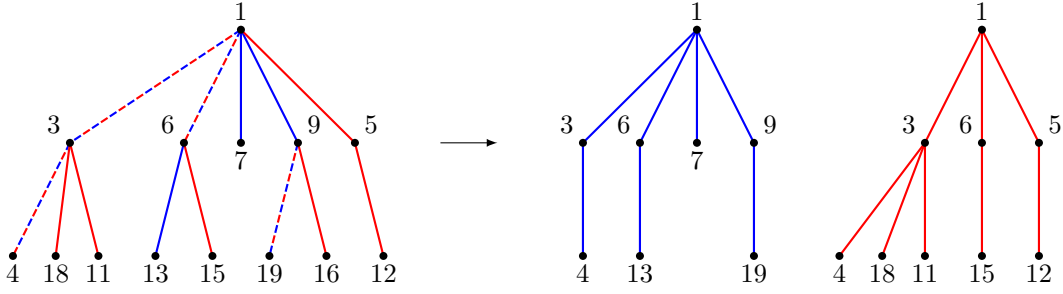

Let us now come back to the correlated graph model, and discuss the local properties of the pair $(\bA,\bC)$, i.e., the aligned graphs before the reshuffling of their vertices through the random permutation. It is convenient to represent this pair of graphs by a single graph $\bG$ whose edge bear a three-valued label represented as a color. Consider that between the vertices $i$ and $j$ there is in $\bG$: 
\begin{itemize}
    \item a blue edge if $\{i,j\}$ is present in $\bA$ but not in $\bC$;
    \item a red edge if $\{i,j\}$ is present in $\bC$ but not in $\bA$;
    \item a bicolored edge if $\{i,j\}$ is present in both $\bA$ and $\bC$;
    \item no edge otherwise.
\end{itemize}
It is clear that this colored graph $\bG$ contains exactly the same information as the pair $(\bA,\bC)$. Suppose now that one performs a local exploration of $\bG$ starting from a vertex $i$, up to a distance $d$, or in other words that one explores simultaneously both graphs $\bA$ and $\bC$ from $i$, following edges that are present in at least one of the two graphs. The arguments invoked to justify the local convergence of a single ER graph can be immediately generalized to this case, and show that with high probability when $n \to \infty$ with $d$ fixed the resulting neighborhood in $\bG$ is a colored GW tree (or multi-type branching process) that we shall denote $\mathcal T$. More precisely, this tree can be built recursively, each vertex having a number of offsprings linked to it by a blue (resp., red, bicolored) drawn as independent Poisson random variables of parameter $\lambda (1-s)$ (resp., $\lambda (1-s)$, $\lambda s$), see the left panel of Fig.~\ref{fig:tree} for an illustration. Thanks to the memoryless property of Poissonian distributions each subtree has the same law as $\mathcal T$.

\subsubsection{Pairs of local neighborhoods with aligned roots}
\label{sec:local_disaligned}

We move now to the pair $(\bA,\bB)$ of disaligned graphs, and consider the following question, whose motivation will be unveiled later on: what is the joint law of $(T,T')$, where $T$ (resp. $T'$) is the depth $d$ neighborhood of a vertex $i$ in the graph $\bA$ (resp. of $i'$ in $\bB$), when $i$ and $i'$ are aligned vertices (i.e. when $i'=\bpis(i)$), with $d$ fixed and $n\to\infty$? This question is obviously related to the exploration process on $\bG$ described above, but with some important differences. Indeed $T$ (resp. $T'$) is built by following only the blue and bicolored (resp., red and bicolored) edges of $\bG$, hence it is obtained from the colored GW tree $\mathcal T$ by keeping only the blue and bicolored (resp. red and bicolored) edges that form the connected component of the root. We can thus define a map $(T,T')=(b({\mathcal T}),r({\mathcal T}))$ that transforms the colored tree $\mathcal T$ into a pair of monochromatic trees $(T,T')$, see the right panel and the caption of Fig.~\ref{fig:tree} for an illustration. In other words once a blue (resp., red) edge has been crossed in the colored GW tree one can draw its descendence as a blue (resp. red) GW tree with a Poisson offspring distribution of parameter $\lambda$. Moreover in the pair $(T,T')$ the original alignment of the vertices is completely lost except for the root vertex. From these observations one can establish a recursive relation for the law of $(T,T')$, to be denoted $\mathbb{P}_1^{(d)}[T,T']$. Denoting $l$ (resp., $l'$) the degree of the root of $T$ (resp., $T'$) and $T_1,\dots,T_l$ (resp., $T'_1,\dots,T'_{l'}$) the subtrees rooted at its descendents, one has
\begin{multline*}
\mathbb{P}_1^{(d)}[T,T']  =
\sum_{l_{\rm b},l_{\rm r},l_{\rm bi}=0}^\infty  \Po(\lambda (1-s);l_{\rm b}) \, \Po(\lambda (1-s);l_{\rm r}) \, \Po(\lambda s;l_{\rm bi}) \, \mathbb{I}(l=l_{\rm b} + l_{\rm bi}) \ \mathbb{I}(l'=l_{\rm r} + l_{\rm bi}) \\
\comprimi\sum_{Q_1,\dots,Q_{l_{\rm b}}} \sum_{R_1,\dots,R_{l_{\rm r}}} \sum_{S_1,S'_1,\dots , S_{l_{\rm bi}},S'_{l_{\rm bi}}} \prod_{i=1}^{l_{\rm b}} \mathbb{P}_0^{(d-1)}[Q_i] \prod_{i=1}^{l_{\rm r}} \mathbb{P}_0^{(d-1)}[R_i] \prod_{i=1}^{l_{\rm bi}} \mathbb{P}_1^{(d-1)}[S_i,S'_i]\\\comprimi\frac{1}{l! l'!}
\sum_{\pi,\pi'} \mathbb{I}((T_1,\dots,T_l)
=\pi(Q_1,\dots,Q_{l_{\rm b}} ,S_1,\dots,S_{l_{\rm bi}})) \, \mathbb{I}((T'_1,\dots,T'_{l'})=\pi'(R_1,\dots,R_{l_{\rm r}},S'_1,\dots,S'_{l_{\rm bi}})) \ ,
\end{multline*}
where $\mathbb{I}(E)$ is the indicator function of the event $E$,
$\pi$ (resp., $\pi'$) is a permutation of its $l$ (resp., $l'$) arguments, and we use the convention $\mathbb{P}_1^{(0)}[T,T']=1$. In this expression $l_{\rm b}$, $l_{\rm r}$ and $l_{\rm bi}$ are the number of blue, red and bicolored edges emerging from the root of the colored GW tree, the $Q_i$'s (resp., $R_i$'s) are the blue (resp., red) usual GW tree rooted at the offsprings reached by a blue (resp., red) edges, and the pairs of trees $(S_i,S'_i)$ are those rooted at offsprings reached by a bicolored edge. The uniform average over the permutations $\pi$ and $\pi'$ arise from the ignorance of the vertex correspondance between the two graphs apart from the aligned root. This expression can be slightly simplified by noting that the relevant information contained in the permutations $\pi$ and $\pi'$ are the indices of the subtrees of $T$ and $T'$ assigned to the correlated pairs of trees $(S_i,S'_i)$. This yields
\begin{align}
\mathbb{P}_1^{(d)}[T,T']  = & \sum_{l_{\rm bi}=0}^{\min(l,l')} \Po(\lambda (1-s);l-l_{\rm bi}) \, \Po(\lambda (1-s);l'-l_{\rm bi})  \, \Po(\lambda s;l_{\rm bi}) \label{eq_P1} \\ &
\frac{1}{\binom{l}{l_{\rm bi}} \binom{l'}{l_{\rm bi}} l_{\rm bi}!}  \sum_{I,I',\sigma} \prod_{i \in I} \mathbb{P}_1^{(d-1)}[T_i,T'_{\sigma(i)}] \prod_{i \in [l] \setminus I} \mathbb{P}_0^{(d-1)}[T_i] \prod_{i \in [l'] \setminus I'} \mathbb{P}_0^{(d-1)}[T'_i] \ , \nonumber
\end{align}
where $I$ (resp $I'$) is a subset of $[l]$ (resp., of $[l']$) of $l_{\rm bi}$ elements, and $\sigma$ a bijection from $I$ to $I'$.

\section{The inference problem}
\label{sec:inference}

\subsection{Estimators}

The inference problem naturally associated with the correlated graph ensemble consists in aligning the graphs $\bA$ and $\bB$, in other words in retrieving the permutation $\bpis$, and therefore the original labeling in $\bC$, from the sole observation of the two graphs $\bA$ and $\bB$. We will study this problem in a Bayesian setting, assuming that the procedure followed for the construction of these graphs is completely known to the observer. All the information available on $\bpis$ given two observed adjacency matrices $A$ and $B$ is thus contained in its posterior probability distribution, that can be expressed thanks to the Bayes formula as
\begin{equation}
\mathbb{P}(\bpis=\pi | \bA=A,\bB=B)= \frac{\mathbb{P}(\bpis=\pi,\bA=A,\bB=B)}{\mathbb{P}(\bA=A,\bB=B)} \propto \mathbb{P}(\bpis=\pi,\bA=A,\bB=B) \ ,
\end{equation}
where here and in the following the symbol $\propto$ implies the presence of a $\pi$-independent prefactor. 

As in all inference problems the notion of optimal estimator depends on the properties required for the estimator, and on the quantitative measure of its distance to the groundtruth signal. In the present context where the groundtruth $\bpis$ is a permutation of $n$ elements, and where the estimator $\pih=\pih(A,B)$ has to be computed from the observed graphs $A$ and $B$, one can envision different possible choices:
\begin{itemize}
    \item If one requires the estimator to be a permutation and if the objective is to minimize the probability that it differs from $\bpis$, then the optimal choice is $\pih(A,B)=\arg\max_\pi\mathbb P(\bpis=\pi | \bA=A,\bB=B)$. As we concentrate in this paper on the constant degree regime where the exact recovery of $\bpis$ is impossible this choice is not relevant here.

\item One can view the groundtruth $\bpis$ as an $n \times n$ matrix with $\{0,1\}$ elements, a 1 in the $i,i'$ entry encoding the fact that $\bpis(i)=i'$. More precisely we can define the ground truth permutation matrix as $\bPis_{ii'}=\mathbb{I}[\bpis(i)=i']$, which is thus constrained to have exactly one 1 per row and per column. In this perspective one can consider an estimator $\bPih$ that is a $\{0,1\}$ matrix, without row and column sum constraints, and measure its accuracy in terms of the Hamming distance between $\bPis$ and $\bPih$ (viewed as strings of $n^2$ bits). The optimal estimator is then
\begin{equation}
    \label{eq_matrix_estimator}
    \bPih(A,B)_{i,i'} = \begin{cases} 1 & \text{if} \quad \mathbb P(\bpis(i) = i' |\bA=A,\bB=B) > \frac{1}{2} \\ 
    0 & \text{if} \quad\mathbb P(\bpis(i) = i' |\bA=A,\bB=B) \le \frac{1}{2}
    \end{cases} \ .
\end{equation}
Note that $P_{i,i'}=\mathbb P(\bpis(i) = i' |\bA=A,\bB=B)$ is a bistochastic matrix, with both row and column sums equal to $1$ (they correspond to sums of probabilities of disjoint events whose union is sure to occur), hence each row and column of $P$ contains at most one element strictly larger than $1/2$. As a consequence the matrix estimator $\bPih$ contains at most one nonzero entry per row and per column, but can leave some row $i$ (or column $i'$) equal to $0$, i.e., do not propose any estimate for the vertex matched to $i$ (or to $i'$). We shall come back on this estimator in \ref{app:matrix_estimator}.

\item The choice on which we will concentrate in most of the following is to require the estimator $\pih = \pih(A,B)$ to be a function from $[n]$ to $[n]$ (not necessarily a permutation), and to measure its quality in terms of the overlap with the groundtruth,
\begin{equation}
\label{eq_overlap}
\ov(\bpih,\bpis) \coloneqq \frac{1}{n} \sum_{i=1}^n \mathbb{I} [\bpih(i) = \bpis(i)] \ ,
\end{equation}
which gives the fraction of vertices in $\bA$ that are correctly assigned their matching vertices in $\bB$. The optimal estimator, in the sense of maximizing this average overlap, is achieved by taking
\begin{equation}
\pih(A,B)(i) = \arg\max_{i'} \mathbb{P}(\bpis(i)=i'|\bA=A,\bB=B) \ .
\label{eq_optimalestimator}
\end{equation}
In intuitive terms this corresponds to compute a $n\times n$ matrix of ``scores'' giving the posterior probability of the event that $i$ was matched to $i'$ given the observations of the graphs $A$ and $B$, and for each $i$ choosing the $i'$ with the highest score.

Two other alternative estimators will be described and evaluated for comparison and complementarity in \ref{app:matrix_estimator}. 

\end{itemize}

\subsection{A local approximation of the posterior}
\label{sec:testhyp}

The optimal estimators discussed above rely on finding the maximum of the posterior probability of $\bpis$, or on computing the probability that the posterior gives to the event $\bpis(i)=i'$. Unfortunately these tasks are computationally intractable: the joint distribution $\mathbb{P}(\bpis=\pi,   \bA=A,\bB=B) $ takes the form  
\begin{multline}
\mathbb{P}(\bpis=\pi,   \bA=A,\bB=B) =\\
=\comprimi  \frac{1}{n!} \prod_{i<j} \left[ 
\left(\frac{\lambda s}{n}\right)^{A_{ij} B_{\pi(i)\pi(j)}} \left(\frac{\lambda}{n}(1-s) \right)^{A_{ij} (1-B_{\pi(i)\pi(j)})+ (1-A_{ij}) B_{\pi(i)\pi(j)}}  \left(1-\frac{\lambda}{n}(2-s) \right)^{(1-A_{ij}) (1-B_{\pi(i)\pi(j)})}
\right]  \\ 
\comprimi = \frac{1}{n!}   \left(1-\frac{\lambda}{n}(2-s) \right)^{\frac{n(n-1)}{2}}
\left(\frac{\frac{\lambda}{n}(1-s) }{1-\frac{\lambda}{n}(2-s)} \right)^{\underset{i<j}{\sum} (A_{ij} +B_{ij})}
\left( \frac{n s}{\lambda (1-s)^2} \left(1-\frac{\lambda}{n}(2-s)\right)\right) ^{\underset{i<j}{\sum} A_{ij} B_{\pi(i)\pi(j)}} \ ,
\end{multline}
hence the posterior probability that is obtained, up to its normalization, by keeping only the terms that depend on $\pi$ in the joint law, reads:
\begin{equation}
\mathbb{P}(\bpis=\pi | \bA=A,\bB=B) \propto \left(\frac{ns}{\lambda (1-s)^2} \left(1-\frac{\lambda}{n}(2-s)\right)\right) ^{\underset{i<j}{\sum} A_{ij} B_{\pi(i)\pi(j)}} \ .
\label{eq_fullposterior}
\end{equation}
Maximising Eq.~\eqref{eq_fullposterior} corresponds thus to solving a quadratic assignment problem, that is notoriously a NP-hard problem (see~\cite{Burkard1998} for a review), and the computation of marginal probabilities of (\ref{eq_fullposterior}) is at least as difficult.

As a consequence we will turn now to approximations of the posterior probability. One possibility, that was investigated in~\cite{BaGlSaWa13,BrBrMaTrWeZe10}, is to write (\ref{eq_fullposterior}) as a factor graph and to derive the Belief Propagation (BP) algorithm associated to it. This is, however, a rather problematic strategy: even if the graphs $\bA$ and $\bB$ are locally tree-like, the constraint that $\pi$ is a permutation has to be implemented by factor nodes with a dense structure and a proliferation of short loops that are quite detrimental for the quality of the BP approximation, unless some side information is provided with the ground-truth values of $\bpis(i)$ revealed for a fraction of vertices $i$. We will follow therefore a different path, exploiting the local tree-like structure of the graphs.

The idea of our computation is to discard a part of the available information and to compute the probability of the event $\bpis(i)=i'$ not under the full posterior given the observation of $\bA$ and $\bB$, but under a truncated posterior where one only observes the local neighborhoods of $i$ and $i'$. To be more precise, let us recall the notation for the truncated matrix $\bA_{i,d}$ (resp.~$\bB_{i',d}$) corresponding to the adjacency matrix of the subgraph of $\bA$ (resp. of $\bB$) induced by the vertices at distance at most $d$ from $i$ (resp. $i'$). The truncated posterior probability of $\bpis(i)=i'$ can be rewritten with Bayes formula as
\begin{align}
\mathbb{P}(\bpis(i)=i'|\bA_{i,d} = T, \bB_{i',d}=T') &= \frac{\mathbb{P}(\bpis(i)=i',\bA_{i,d}=T, \bB_{i',d} = T') }{\mathbb{P}(\bA_{i,d} = T, \bB_{i',d} = T')} \nonumber \\
&= \frac{\mathbb{P}(\bA_{i,d} = T, \bB_{i',d} =T' | \bpis(i)=i') \mathbb{P}(\bpis(i)=i') }{\mathbb{P}(\bA_{i,d} = T, \bB_{i',d} = T')} \nonumber \\
& = \frac{1}{n} \frac{\mathbb{P}(\bA_{i,d} = T, \bB_{i',d} =T'|\bpis(i)=i')}{\mathbb{P}(\bA_{i,d} = T, \bB_{i',d} = T')} \ ,
\label{eq_truncated_posterior}
\end{align}
where in the last line we used the fact the prior probability of the event $\bpis(i)=i'$ is $1/n$. We claim that in the large $n$ limit with $d$ fixed the observed neighborhoods $T$ and $T'$ are trees with high probability, and that the fraction in (\ref{eq_truncated_posterior}) converges to the ratio of the probabilities introduced in the local properties of the random graphs in Section \ref{sec:local_props}, namely
\begin{equation}
\lim_{n \to \infty}\frac{\mathbb{P}(\bA_{i,d} = T, \bB_{i',d} =T'|\bpis(i)=i')}{\mathbb{P}(\bA_{i,d} = T, \bB_{i',d} = T')} = \frac{\mathbb P_1^{(d)}[T,T']}{\mathbb P_0^{(d)}[T]\mathbb P_0^{(d)}[T']} \ ,
\label{eq_ratio}
\end{equation}
with $\mathbb P_0^{(d)}$ and $\mathbb P_1^{(d)}$ defined in equations (\ref{eq_P0}) and (\ref{eq_P1}) respectively. As a matter of fact the numerator of the left hand side is precisely the quantity we investigated in Section \ref{sec:local_disaligned}, and in the denominator the permutation $\bpis$ is averaged out, hence the vertex $i'$ can be seen as uniformly chosen in the aligned graph $\bC$. With a probability $1-O(1/n)$ the depth $d$ neighborhoods of $i$ and $i'$ do not intersect in $\bG$, hence $T$ and $T'$ are asymptotically independent and drawn from their marginal probabilities. From the expressions given in equations (\ref{eq_P0}) and (\ref{eq_P1}) we can derive a recursive expression of this ratio, that we will denote $L^{(d)}(T,T')$. After some simplifications of the Poisson probabilities one obtains indeed
\begin{equation}
L^{(d)}(T,T') = \frac{\mathbb P_1^{(d)}[T,T']}{\mathbb P_0^{(d)}[T]\mathbb P_0^{(d)}[T']} = 
\sum_{l_{\rm bi}=0}^{\min(l,l')} \e^{\lambda s} (1-s)^{l+l'}\left( \frac{s}{\lambda(1-s)^2} \right)^{l_{\rm bi}} 
 \sum_{I,I',\sigma} \prod_{i \in I} L^{(d-1)}(T_i,T'_{\sigma(i)}) \ , 
\label{eq_Ld}
\end{equation}
with $L^{(0)}(T,T')=1$, and where we recall that as in equation (\ref{eq_P1}) $T_1,\dots,T_l$ are the subtrees of $T$ rooted at the $l$ offsprings of the root of $T$, $T'_1,\dots,T'_{l'}$ is a similar decomposition of $T'$, $I$ (resp.~$I'$) is a subset of $[l]$ (resp. of $[l']$) of $l_{\rm bi}$ elements, and $\sigma$ a bijection from $I$ to $I'$.

Note that $L^{(d)}$ is the ratio of the probabilities for the generation of the pair $(T,T')$ in two different ensembles, a correlated one with $\mathbb{P}_1^{(d)}$ and an uncorrelated one with the product of the $\mathbb{P}_0^{(d)}$. According to the Neyman-Pearson lemma this likelihood ratio can be used to design the optimal estimators for the hypothesis testing problem where an observer is handed a sample $(T,T')$ and must decide whether the pair was generated in a correlated way from $\mathbb{P}_1^{(d)}$ (alternative hypothesis) or as a pair of independent copies from the marginal law $\mathbb{P}_0^{(d)}$ (null hypothesis). The optimal answer to this question is the alternative if and only if $L^{(d)}(T,T') \ge \alpha$, where $\alpha$ is a threshold that depends on the compromise to be made between false positive and false negative errors. In~\cite{GaMaLe21b} this hypothesis testing problem has been studied per se, and translated to the graph alignment problem, here we arrived at it through a slightly different road, namely the study of the truncated posterior. 

To simplify notations in the following we abstract Eq.~(\ref{eq_Ld}) by introducing a function $f$ that takes as inputs two integers $l$ and $l'$ and an $l \times l'$ array $L_{i,i'}$ of real numbers and computes
\begin{equation}
f(l,l';\{L_{i,i'}\})=
\sum_{l_{\rm bi}=0}^{\min(l,l')} \e^{\lambda s} (1-s)^{l+l'}\left( \frac{s}{\lambda(1-s)^2} \right)^{l_{\rm bi}} 
 \sum_{I,I',\sigma} \prod_{i \in I} L_{i,\sigma(i)} \ , 
\label{eq_function_f}
\end{equation}
with the same definitions for $I$, $I'$ and $\sigma$ as in (\ref{eq_Ld}), and with the convention $f=\e^{\lambda s} (1-s)^{l+l'}$ if $\min(l,l')=0$. In the limit of perfect correlation $s\to 1$ this function becomes
\begin{equation}
f(l,l';\{L_{i,i'}\})= \mathbb{I}(l=l') \e^{\lambda} \lambda^{-l}  
 \sum_{\sigma} \prod_{i=1}^l L_{i,\sigma(i)} \ .
\end{equation}

\subsection{A message passing algorithm for the graph alignment problem}

We shall now present the algorithm for the alignment of correlated Erd\H os-R\'enyi random graphs that follows naturally from the above considerations. Given the observations of two graphs $A$ and $B$ and a positive integer parameter $d$  we compute an estimator $\pih=\pih(A,B)$ which is a function (not necessarily bijective) of the vertex set of $A$ to the one of $B$, with the goal to maximize the average overlap of this correspondence with the groundtruth. As the information theoretical optimal procedure given in (\ref{eq_optimalestimator}) is not computationally tractable we approximate the posterior distribution by its truncated version defined in (\ref{eq_truncated_posterior}), which corresponds to define a $n \times n$ score matrix $L_{ii'}^{(d)}=L^{(d)}(A_{i,d},B_{i',d})$ and set $\pih(i)=\arg\max_{i'} L_{ii'}^{(d)}$, the constant multiplicative prefactor $1/n$ in (\ref{eq_truncated_posterior}) being irrelevant here. Thanks to the recursive nature of $L^{(d)}$ these scores can be computed via a message-passing procedure; to specify it we need first to introduce some additional notations. For a vertex $i$ of $A$ we write $\partial i$ for the set of vertices adjacent to $i$, and $d_i=|\partial i|$. Similar notations apply to vertices $i'$ of $B$; to avoid any confusion we will always use primed indices for vertices in $B$, and consider the vertex set $\{1,\dots,n\}$ of $A$ as distinct from $\{1',\dots,n'\}$, the vertex set of $B$. We now introduce a set of messages $L_{ii' \to jj'}^{(t)}$ for all vertices $i$ of $A$, all vertices $i'$ of $B$, all $j \in \partial i$, and all $j' \in \partial i'$. The discrete `time' index $t$ corresponds to a number of iterations, or depth; the interpretation of $L_{ii' \to jj'}^{(t)}$ is the likelihood ratio for the depth $t$ neighborhoods of $i$ and $i'$, deprived of the branch linking them to $j$ and $j'$ respectively. We can thus summarize our algorithm as follows:
\begin{enumerate}
\item all messages are initialized to $L_{ii' \to jj'}^{(0)}=1$.
\item for each $t=1,\dots,d-1$, they are updated according to
\begin{equation}
L_{ii' \to jj'}^{(t)} = f(d_i-1,d_{i'}-1;\{L_{kk' \to ii'}^{(t-1)} \colon k \in \partial i \setminus j \, , \,  k' \in \partial i' \setminus j' \} ) \ ,
\label{eq_MP}
\end{equation}
where the function $f$ was defined in Eq.~(\ref{eq_function_f}).
\item the scores are computed as
\begin{equation}
L_{ii'}^{(d)} = f(d_i,d_{i'};\{L_{jj' \to ii'}^{(d-1)}  \colon j \in \partial i \, , \,  j' \in \partial i' \} ) \ .
\end{equation}
\item finally the estimator $\pih$ is computed as $\pih(i)=\arg\max_{i'} L^{(d)}_{ii'}$, with ties broken uniformly at random if several $i'$ achieve the same maximal score.
\end{enumerate}

Let us make a series of comments before presenting the numerical results we obtained with this algorithm: 
\begin{itemize}
    \item Cycles certainly occur in random graphs, that are only locally tree-like; one may thus wonder about the meaning of $L_{ii'}^{(d)}$ when the depth $d$ neighborhoods of $i$ and $i'$ are not both trees. A moment of thought reveals that the iterative procedure described in the algorithm actually computes the likelihood ratio of the trees of non-backtracking walks of length at most $d$ starting at $i$ and $i'$ (also known as the computational tree). This coincides with the usual neighborhood when the latter is a tree, and otherwise ``unwraps'' the cycles according to this non-backtracking rule. Note also that there is only a finite number (on average) of cycles of finite length in random graphs with fixed average degrees. The neighborhood of depth $d$ of most vertices is acyclic even if $d$ grows (logarithmically) with $n$, see for instance~\cite{GaMa20} for a precise statement of the coupling between random graphs and random trees on logarithmic scales. 

\item The interpretation put forward in~\cite{GaMaLe21b} of the scores as likelihood ratios for an hypothesis testing problem between two perfectly aligned versus two independent trees may suggest an issue when using these scores $L^{(d)}_{ii'}$ between partially correlated trees, i.e. when $i'\neq \bpis(i)$ but with $i$ and $\bpis^{-1}(i')$ at a graph distance smaller than $2d$ in $\bG$, since in that case the joint law of the neighborhoods is neither the null nor the alternative of the hypothesis testing problem. We emphasize however that our derivation of the algorithm follows from the truncation of the posterior probability of the hidden permutation, and does not rely on the hypothesis testing interpretation. Moreover we have checked that this potential issue does not spoil the behavior of the algorithm, the scores between partially aligned neighborhoods being substantially smaller than between perfectly aligned ones, see \ref{app:scores} for more details on this point.

\item The number of messages $L_{ii' \to jj'}$ is $4$ times the product of the number of edges in the two graphs, i.e., $O(n^2)$ in the sparse regime considered here. The algorithm requires therefore a total number of message updates of order $O(d \, n^2)$. However, the number of operations required for the computation of the function $f$ in (\ref{eq_function_f}) grows very fast (factorially) with $\min\{l,l'\}$, because of the sum over the permutations $\sigma$. As the maximal degree in an ER graph grows slowly with $n$, as $O\left(\frac{\ln n}{\ln \ln n}\right)$, the asymptotic scaling with $n$ of the computational cost of the algorithm remains polynomial in $n$. Nevertheless from a practical point of view this factorial growth shall restrict our study to random graphs with rather small average degree (for instance if $l=l'=10$ the computation of $f$ would involve a sum over more than $10^8$ terms).

\item Another implementation remark concerns the very large value some messages can acquire for growing $d$; to alleviate this problem we actually stored the logarithms of the messages.

\item An implementation of the algorithm in C language can be found at \url{https://github.com/giovannipiccioli/graph_alignment}.

\end{itemize}

\section{Numerical results for the graph alignment algorithm}
\label{sec:numerical}

This Section shall be devoted to a presentation of the results we have obtained by numerical simulations of the algorithm presented above. Let us recall that the parameters of the problem are $n$, the number of vertices of the pair of graphs to be aligned, $\lambda$, their average degree, $s$, their correlation, and $d$, the parameter of the algorithm that controls the depth at which the neighborhoods of the vertices are inspected to decide which pairs of nodes to match. For each choice of these parameters we compute $\mathbb{E}[\ov(\bpih,\bpis)]$, where the overlap between the groundtruth permutation $\bpis$ and the estimate returned by the algorithm $\bpih$ has been defined in Eq.~(\ref{eq_overlap}), and the expectation is over the randomness in the generation of the pair $(\bA,\bB)$ of graphs, and possibly over the tie-breaking procedure of the algorithm; to simplify the notation we keep implicit the dependency of this average overlap upon the parameters $(n,\lambda,s,d)$. We estimate numerically this expectation by an empirical average over several independent samples. Our ultimate goal would be to determine the algorithmic phase diagram in the $(\lambda,s)$ plane in the large size limit $n\to \infty$, in other words to determine the values of these parameters for which the algorithm achieves asymptotically a partial recovery of the hidden permutation; in formula this corresponds to $\underset{n\to\infty}{\liminf} \, \mathbb{E}[\ov(\bpih,\bpis)] > 0$, for a suitable choice of the depth parameter $d$, possibly $n$-dependent. We expect this property to be monotonous in $s$, as increasing the correlation between the two graphs increase the amount of information available for the inference of $\bpis$, we would like thus to determine the algorithmic threshold $s_{\rm algo}(\lambda)$ defined as the smallest value of $s$ for which the algorithm achieves asymptotically partial recovery for the parameters $(\lambda,s)$. One knows from previous works that partial recovery is only achievable in some regions of the $(\lambda,s)$ plane; in particular it has been proven in~\cite{GaMaLe21a} that the asymptotic average overlap of any estimator (be it efficiently computable or not) is upper bounded by $c(\lambda s)$, the largest non-negative solution of $1-c=\e^{-\lambda s c}$, which corresponds to the fraction of vertices in the largest component of the intersection graph between $\bA$ and $\bC$, an Erd\H os-R\'enyi random graph of average degree $\lambda s$. As $c(\lambda s)=0$ whenever $\lambda s \le 1$, this implies that partial recovery is impossible in this case, hence the lower bound on the algorithmic threshold $s_{\rm algo}(\lambda) > 1/\lambda$. As we shall see the determination of the algorithmic phase transition is a very challenging numerical task, we have nevertheless some partial answers to this question.

\subsection{The scaling of the depth parameter $d$}

\begin{figure}
    \centering
    \includegraphics{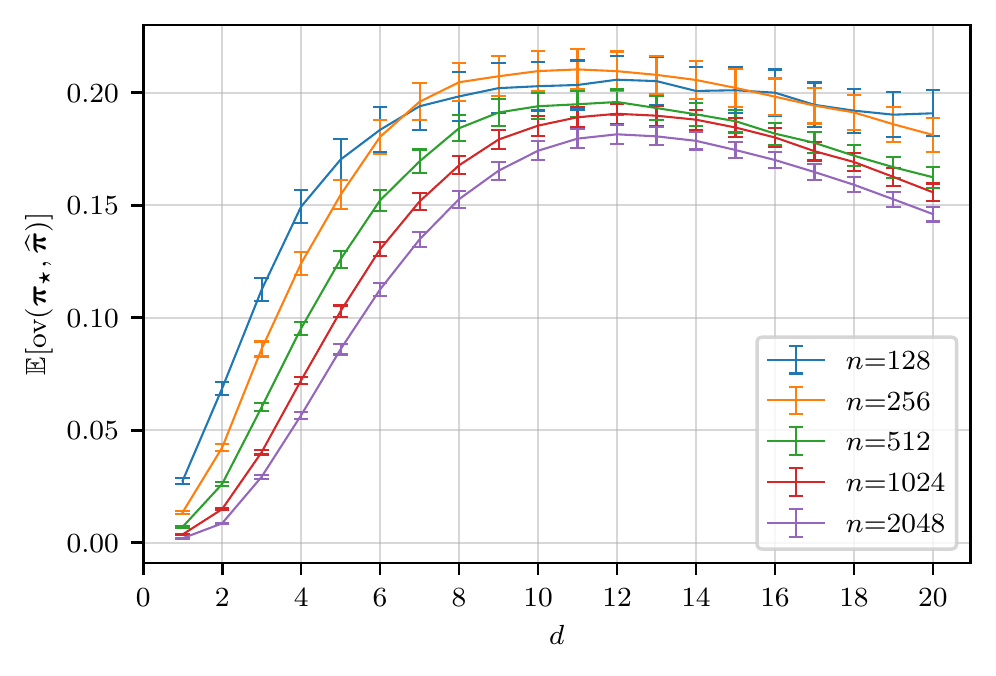}
    \caption{The average overlap as a function of the depth $d$ for average degree $\lambda=1.4$, correlation $s=0.81$, and various values of system size $n$. Each point is averaged over $100$ realizations of the two graphs $\bA$, $\bB$. The overlap between the estimator $\bpih$ and the ground truth $\bpis$ exhibits a clear maximum for an intermediate value of the depth $d$, which increases slowly with the system size $n$.}
    \label{fig:overlap_vs_d}
\end{figure}

Let us first discuss the choice of the depth parameter $d$ of the algorithm, that we have left unspecified up to now. On the one hand we would like to take $d$ as large as possible: the algorithm has been derived by replacing the full posterior distribution (\ref{eq_fullposterior}), that exploits all the information contained in the realization of $(\bA,\bB)$, by its truncated version (\ref{eq_truncated_posterior}) which only depends on $(\bA_{\cdot,d},\bB_{\cdot,d})$, thus discarding a part of the available information. Larger values of $d$ corresponds to a less drastic loss of information, the truncated posterior getting closer to the information-theoretical optimal full posterior. On the other hand, for a finite value of $n$ we should not take $d$ arbitrarily large: the computation that underlies the algorithm is based on the assumption that the neighborhoods explored up to depth $d$ are trees, which is not true when $d$ exceeds $O(\ln n)$, the scaling of the minimal length of a cycle from a typical vertex in a sparse random graph. When a cycle is encountered the algorithm unwraps it according to the non-backtracking rule, hence producing spurious terms that corrupt the estimator. As a consequence one expects that for a given choice of the parameters $(n,\lambda,s)$ there will be an optimal value of $d$ that reaches a compromise between these two conflicting requirements, for which the average overlap will be maximal. This is confirmed by the results presented in Fig.~\ref{fig:overlap_vs_d}, which shows the average overlap as a function of $d$, for one choice of $(\lambda,s)$ and several values of $n$: these curves exhibit indeed a maximum at an optimal value $d_*(n,\lambda,s)$.

In order to make statements about the asymptotic behavior of the algorithm in the limit $n \to \infty$ one should now understand what is the scaling of the optimal depth $d_*(n,\lambda,s)$ in this limit. For the range of values of $n$ displayed in Fig.~\ref{fig:overlap_vs_d} one sees a mild increase of $d_*$ with $n$, one could thus be tempted to assume that it reaches a finite value when $n$ diverges. A moment of thought reveals that this expectation is wrong, and, as a matter of fact, for any value of $(\lambda,s)$ the average overlap is at most of order $1/n$, and hence vanishes, if the large $n$ limit is taken with any fixed finite value $d$. To justify this claim let us consider a vertex $i$ in $\bA$, its image $i'=\bpis(i)$ in $\bB$ through the ground-truth permutation, and let us denote $T_i$ and $T'_{i'}$ their respective neighborhoods of depth $d$ in the two graphs. There are two failure mechanisms in the algorithm that lead to a wrong estimation of the vertex matched to $i$, i.e. to $\pih(i)\neq i'$: either a vertex $j' \neq i'$ achieves a strictly higher score than the correct assignment, $L_{i,j'}^{(d)} > L_{i,i'}^{(d)}$, or $L_{i,i'}^{(d)}$ achieves the maximum score in $L_{i,\cdot}^{(d)}$ but $i'$ is not the unique maximizer. In the latter case the probability that $\pih(i)=i'$ is one divided by the degeneracy of the maximum, as the algorithm picks a uniformly random maximizer of the score in case of ties. It turns out that if $n\to \infty$ with $d$ fixed the number of vertices $j'$ that achieves the same score as the correct match $i'$ is extensive (proportional to $n$): $T'_{i'}$ being a fixed finite tree, the probability $\mathbb{P}_0^{(d)}[T'_{i'}]$ that the depth $d$ neighborhood of a randomly chosen vertex $j'$ is isomorphic to it is of order 1, hence the cardinality of $\{j' \ : \ L_{i,j'}^{(d)} = L_{i,i'}^{(d)} \}$ is of order $n$. This concludes the justification of our claim that the average overlap is at most of order $1/n$ in the large $n$ limit with $d$ fixed: even if $i'$ achieves the maximum in $L_{i,\cdot}^{(d)}$ there will be an extensive number of vertices achieving it, hence $i'$ will be picked with a probability inversely proportional to this degeneracy.

\begin{figure}
    \centering
    \includegraphics[width=\columnwidth]{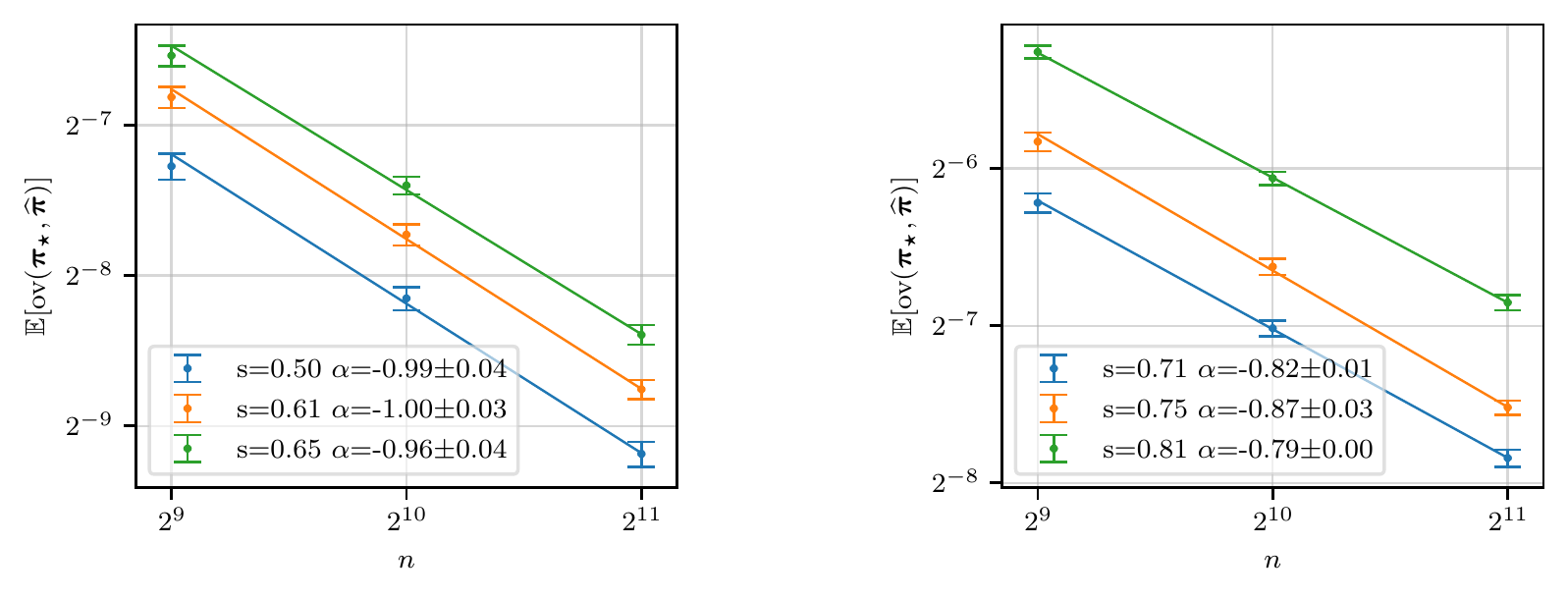}
    \caption{The average overlap between the estimated and ground truth configurations $\mathbb{E}[\ov(\bpih,\bpis)]$ as a function of the system size $n$ for a fixed depth $d=2$, average degree $\lambda=1.4$ and for the left panel correlations $s=0.5$, $s=0.61$, $s=0.65$, for the right panel $s=0.71$, $s=0.75$, $s=0.81$. Every point is an average over $100$ samples. The lines correspond to power-laws fits of the form $\mathbb{E}[\ov(\bpih,\bpis)]= c \, n^{\alpha}$ with $c$ and $\alpha$ two fitting constants, logarithmic scales being used on both axis. In the left panel the exponent $\alpha$ is compatible with the value $-1$ of the analytic argument, while on the right $\alpha>-1$, which reveals strong finite size effects and a preasymptotic behavior.}
    \label{fig:ovlap_small_d}
\end{figure}

This reasoning is confirmed by the numerical results presented in Fig.~\ref{fig:ovlap_small_d}, where we plot the average overlap as a function of $n$ for a fixed (and small) value of the depth, $d=2$. The left panel, for some choices of $(\lambda,s)$, shows indeed a behavior of the form $\mathbb{E}[\ov(\bpis,\bpih)]\propto n^\alpha$ with $\alpha=-1$. The right panel, for slightly different parameters $(\lambda,s)$, exhibit a power-law behavior with $\alpha > -1$; we interpret this apparent contradiction with the reasoning above as signalling that the values of $n$ investigated (which are the largest ones we could reach within a reasonable amount of computation time) are too small to be representative of the asymptotic behavior of the limit $n \to \infty$. Indeed the probability $\mathbb{P}_0^{(d)}[T'_{i'}]$ is certainly of order 1, but can be numerically very small; as long as $n \mathbb{P}_0^{(d)}[T'_{i'}] \ll 1$ the typical number of confounding vertices is on average much smaller than 1, and hence typically 0.

Having ruled out the possibility that the optimal depth $d_*(n,\lambda,s)$ remains constant as $n\to\infty$ leaves open the question of its scaling with $n$ in this limit. We conjecture that $d_*(n,\lambda,s)= \widetilde{d}_*(\lambda,s) \ln n + o(\ln n)$, with a constant $\widetilde{d}_*$ possibly dependent on $(\lambda,s)$. Indeed $d=\Theta(\ln n)$ is the largest possible scaling of $d$ that ensures that most of the depth-$d$ neighborhoods of the graph are trees, compromising between the two opposed requirements on $d$ discussed above. Moreover with this scaling of $d$ the typical degeneracy $n \mathbb{P}_0^{(d)}[T'_{i'}]$ of the scores remains bounded, invalidating the previous ``pigeonhole'' argument used in the limit $n \to \infty$ with $d$ fixed, and opens the possibility of asymptotic partial recovery in some regions of the $(\lambda,s)$ plane. Unfortunately it does not seem possible to test numerically this conjecture: on the range of $n$ that is accessible to our simulations the function $\ln n$ has very small variations that do not allow for an accurate fit.

Another element of information on the values of depth $d$ for which the truncated posterior behaves as expected in Bayes-optimal inference is presented in \ref{app:nishimori} where we test the validity of the so-called Nishimori condition for the truncated estimator. We observe that the Nishimori condition is violated for larger values of the depth $d$. This implies that the presented message passing algorithm will be suboptimal for those values of $d$.

\subsection{The performances of the algorithm across the $(\lambda, s)$ plane}
 
\begin{figure}
    \centering
    \includegraphics[width=0.75\columnwidth]{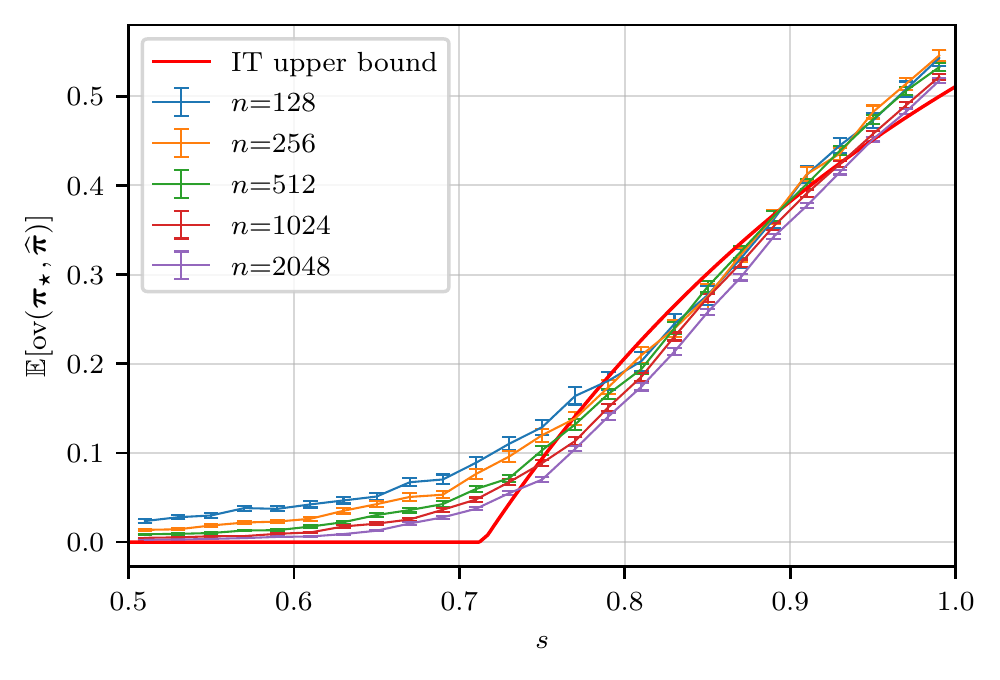}
    \caption{The average overlap $\mathbb{E}[\ov(\bpih,\bpis)]$ as a function of the correlation $s$ for average degree $\lambda=1.4$, depth $d=10$ and several values of system sizes $n$. The curve denoted `IT upper bound' refers to the upper bound of~\cite{GaMaLe21a} on the overlap given by the fraction of vertices in the giant component of an Erd\H os-R\'enyi random graph of average degree $\lambda s$.}
    \label{fig:ovlap_fixed_d10}
\end{figure}

One can adopt two attitudes in facing this difficulty in the choice of $d$: if one is interested in the performances reachable in practice by the algorithm, then one has only to consider relatively small values of $n$, at most a few thousands, for which $d_*$ varies only mildly, and study the overlap for a fixed, reasonably large value of $d$. This is what we have done to produce the curves of Fig.~\ref{fig:ovlap_fixed_d10}, choosing here $d=10$. As anticipated above these curves are increasing functions of $s$, confirming that more correlated graphs are easier to align. For some values of the parameters we obtain average overlaps which are higher than the upper bound derived in~\cite{GaMaLe21a} in terms of the fraction of vertices in the giant component of the bicolored subgraph of $\bG$, an Erd\H os-R\'enyi random graph of average degree $\lambda s$. This of course is not a contradiction, the bound is valid asymptotically in the $n\to \infty$ limit while our numerical results are obtained at finite $n$, for which the size of the largest components of the bicolored subgraph of $\bG$ have strong fluctuations (it would thus be useful to derive an information-theoretic upper bound on the average overlap valid for all $n$, or at least compute the finite size corrections to the asymptotic one). It shows nevertheless that the sizes of a few thousands that one can reach numerically suffer from strong finite size effects and are still far from the asymptotic behavior; as a matter of fact we argued above that in the limit $n\to\infty$ taken with $d$ fixed (which is the case on Fig.~\ref{fig:ovlap_fixed_d10}) the average overlap vanishes, a fact which is certainly impossible to deduce from a naive extrapolation of the data in Fig.~\ref{fig:ovlap_fixed_d10}.

\begin{figure}
    \centering
    \includegraphics[width=\columnwidth]{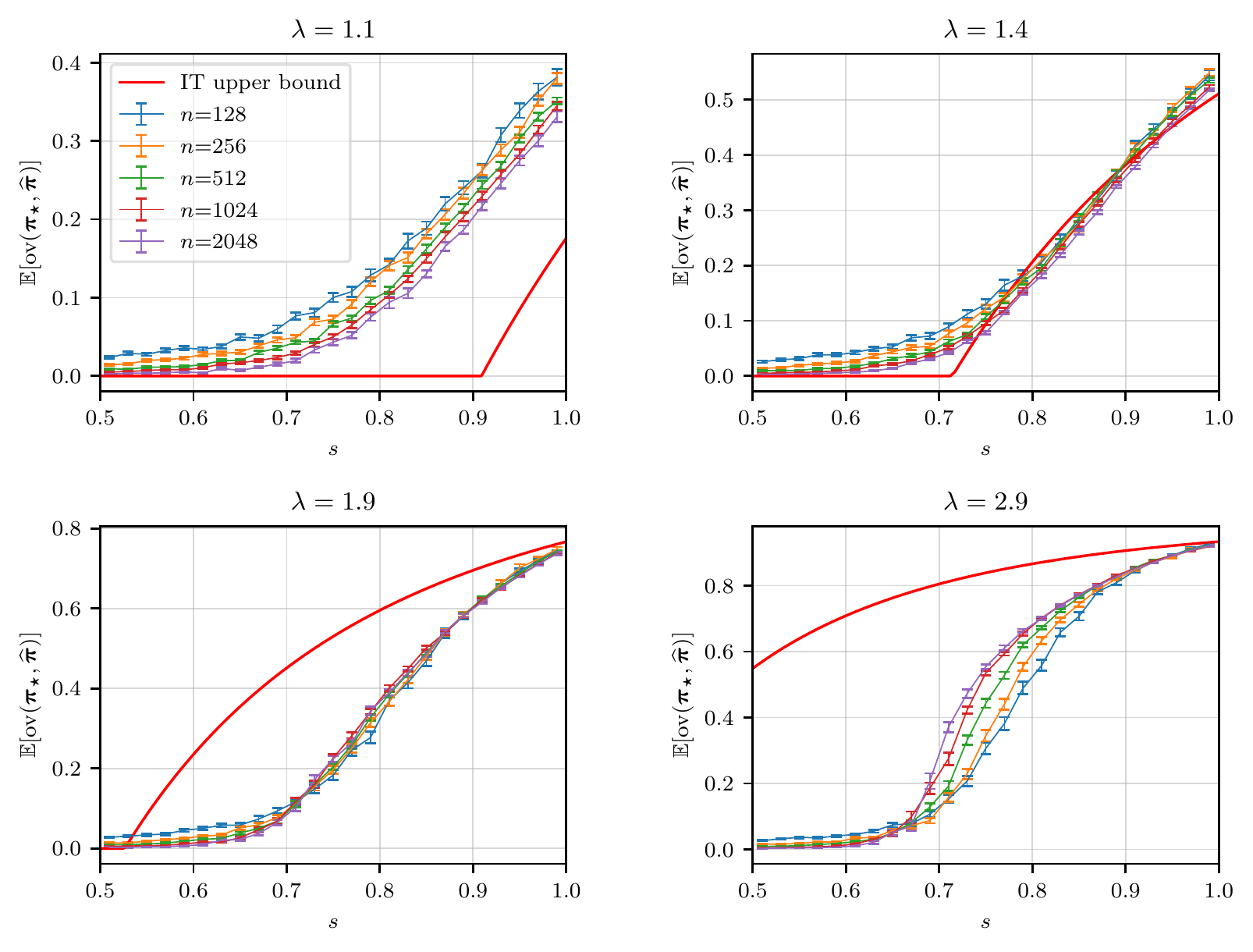}
    \caption{The average overlap $\mathbb{E}[\ov(\bpih,\bpis)]$ computed at the optimal depth, as a function of the correlation $s$, with average degree $\lambda=1.1$ (top left panel), $\lambda=1.4$ (top right panel), $\lambda=1.9$ (bottom left panel), $\lambda=2.9$ (bottom right panel), for several values of system size $n$, the color code being the same in all panels.}
    \label{fig:optimal_ovlap}
\end{figure}

\begin{figure}
    \centering
    \includegraphics[width=\columnwidth]{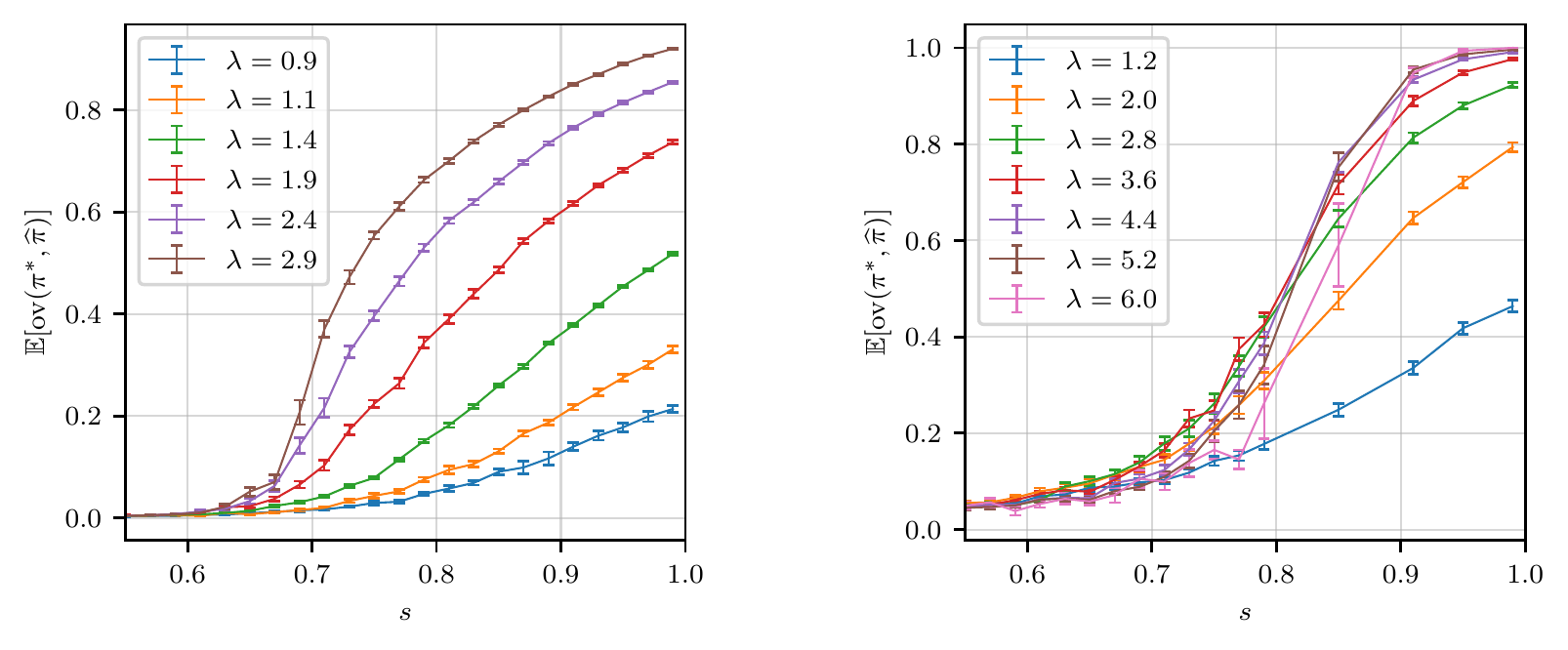}
    \caption{The average overlap $\mathbb{E}[\ov(\bpih,\bpis)]$ computed at the optimal depth as a function of the correlation $s$ for several values of the average degree $\lambda$. (Left) Results obtained for system size $n=2048$: all the curves are zero below $s\approx 0.6$. For $s>0.6$, the overlap increases faster in curves with greater values of $\lambda$.
    (Right) Results obtained for $n=64$: we consider higher values of $\lambda$, up to $\lambda=6$. When $s$ tends to $1$ curves with higher $\lambda$ converge to higher values of the overlap.
    }
    \label{fig:overlap_vs_s_N2048_many_lambdas}
\end{figure}

If on the other hand one is interested in making conjectures on the $n\to \infty$ limit behavior of the algorithm from the finite $n$ results, one needs to find a meaningful way to extrapolate them, taking into account the necessary dependency of $d$ on $n$. To avoid the problematic choice of the prefactor in front of $\ln n$ in $d$ we adopted the following pragmatic strategy: for every value of $(n,\lambda,s)$ investigated we computed the average overlap for all values of $d$ with $1\leq d\leq20$, and selected the one that maximized the average overlap, defined above as $d_*(n,\lambda,s)$. This eliminates one parameter and in principle allows for an extrapolation at large $n$ for fixed $(\lambda,s)$; the results presented in figures~\ref{fig:optimal_ovlap} and \ref{fig:overlap_vs_s_N2048_many_lambdas} have been obtained by following this procedure.
 
\begin{figure}
    \centering
    \includegraphics[width=\columnwidth]{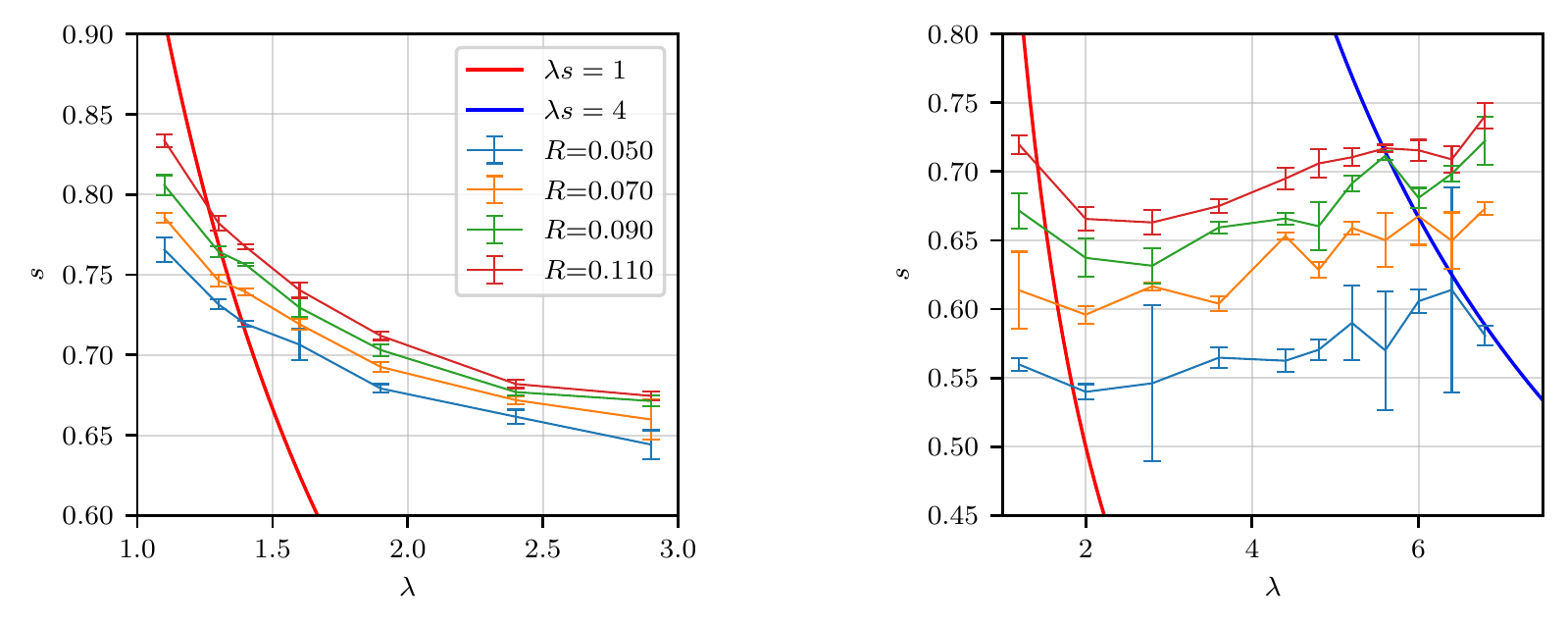}
    \caption{The crossover lines in the $(\lambda,s)$ plane, defined as the minimal value of the correlation $s$ for which the overlap exceeds a small threshold $R$. (Left) $n=2048$. (Right) $n=64$: the crossover position is approximately constant for large values of $\lambda$. We compare these lines with the lower and upper bounds on the information theoretic threshold for the possibility of asymptotic partial recovery, namely $\lambda s = 1$ and $\lambda s = 4$. }
    \label{fig:crossover_diagram}
\end{figure}

The top left panel of figure~\ref{fig:optimal_ovlap}, corresponding to $\lambda=1.1$, shows average overlaps exceeding considerably the upper bound of~\cite{GaMaLe21a}, even in the regime $\lambda s <1$ where the latter vanishes (the top right panel for $\lambda=1.4$ displays a similar but less marked phenomenon): effective finite-size partial recovery is achievable even when the asymptotic one is impossible. As explained above this is a finite size effect, with the gap between the numerical results and the $n\to\infty$ bound closing at a very slow rate of order $1/\ln(n)$ (doubling $n$ produces a constant downward shift in the overlap curves for the ranges investigated). On the contrary the average overlaps in the bottom panels of figure~\ref{fig:optimal_ovlap}, corresponding to $\lambda=1.9$ and $\lambda=2.9$, are well below the upper bound, and suggest that the algorithmic threshold $s_{\rm algo}(\lambda)$ is markedly above the lower bound $1/\lambda$. In principle the determination of $s_{\rm algo}(\lambda)$ should simply follow from an extrapolation of these curves in the $n\to\infty$ limit, the extrapolation being 0 for $s<s_{\rm algo}(\lambda)$ and strictly positive for $s > s_{\rm algo}(\lambda)$; unfortunately, because of the relatively small sizes we could reach, of the slowly vanishing finite-size effects and in absence of additional analytic arguments to constrain the fitting form we did not manage to find a reliable and stable extrapolation procedure. Nevertheless it is  tempting from a visual inspection of the bottom panel of Fig.~\ref{fig:optimal_ovlap} to propose an estimation of $s_{\rm algo}(\lambda)$ as the value of $s$ for which the average overlap markedly grows away from 0, for the largest available size (here $n=2048$). For both these two values of $\lambda$ this rough estimate is slightly above $s \approx 0.6$; this very weak dependency on $\lambda$ is confirmed by the data presented in the left panel of Fig.~\ref{fig:overlap_vs_s_N2048_many_lambdas}, which shows on the same plot various values of $\lambda$ for $n=2048$. All the curves collapse to zero for $s<0.6$, indicating that the algorithm does not achieve partial recovery in this regime. Even if one does not see a sharp transition on these finite-size results all the curves seem to start to grow above roughly the same value $s \approx 0.6$. To confirm this finding we performed some numerical experiments for larger values of $\lambda$, up to $\lambda=6.8$, for which we were limited to much smaller sizes ($n=64$) because of the factorial growth of the computational cost with the degree of the vertices; these results, shown in the right panel of Fig.~\ref{fig:overlap_vs_s_N2048_many_lambdas}, exhibit also a notable growth of the average overlap around $s \approx 0.6$. To be slightly more quantitative we introduce a small arbitrary threshold $R$ and define an effective transition point (or more precisely a crossover) as the minimal value of $s$ for which the overlap is larger than $R$, which thus depends on $R$, $n$ and $\lambda$. These effective transition lines in the $(\lambda,s)$ phase diagram are presented in Fig.~\ref{fig:crossover_diagram} for two values of $n$ (2048 in the left panel and 64 in the right one) and several values of $R$. Once again the finite size effects lead to partial recovery in the information theoretic impossible regime. Moreover for $n=64$ and large values of $\lambda$ one observes that the crossover line enters the information theoretic feasible phase, thus hinting at the existence of a hard phase, i.e. a phase where the problem is information theoretically feasible but the algorithm fails to perform partial recovery.

The conclusion of this Section is that the numerical experiments show for practical sizes of the order $n\approx 10^3$  and $0.9\leq\lambda\leq2.9$ an observed algorithmic threshold for partial recovery almost independent on $\lambda$, around $s \approx 0.6$. 

\section{The tree problem}
\label{sec:tree}

We have seen in the previous Section that the determination of the algorithmic threshold $s_{\rm algo}(\lambda)$ from the finite $n$ numerical simulations on pairs of graphs suffered from strong finite $n$ corrections that limited its accuracy. In this Section we shall follow a different road, that amounts in some sense to work directly with $n=\infty$; nevertheless we will have to face some other numerical difficulties, related to finite $d$ effects.

\begin{figure}
    \centering
    \includegraphics[width=10cm]{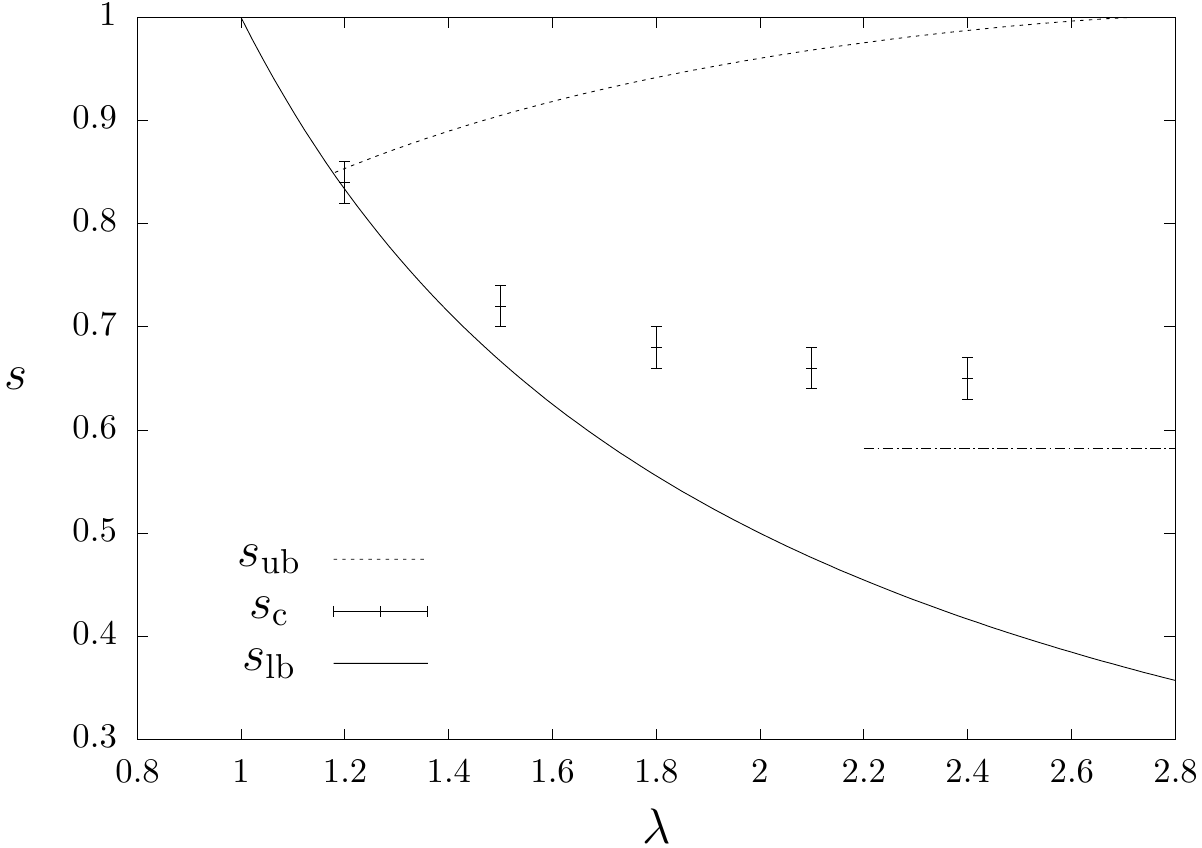}
    \caption{The numerically determined threshold $s_{\rm c}(\lambda)$ for the hypothesis testing problem on random trees, along with its lower bound $s_{\rm lb}(\lambda)=1/\lambda$ and its upper bound $s_{\rm ub}(\lambda)$ that follows from theorem 3 in~\cite{GaMaLe21b}. The meaning of the horizontal dot-dashed line is explained in the conclusions.}
    \label{fig:tree_pd}
\end{figure}

The computation of the probability of the event $\bpis(i)=i'$ under the truncated posterior made naturally appear the ratio $L^{(d)}(T,T') = \frac{\mathbb P_1^{(d)}[T,T']}{\mathbb P_0^{(d)}[T]\mathbb P_0^{(d)}[T']}$ of probabilities for the generation of a pair of trees in a correlated and uncorrelated ensemble of random trees. According to the Neyman-Pearson lemma this likelihood ratio leads, by thresholding, to the family of optimal estimators in the hypothesis testing problem where an observer has to decide from which of these two laws a pair of trees has been generated. Forgetting temporarily the original graph alignment problem we concentrate for the moment on this hypothesis testing problem on trees (note that the parameter $n$ does not appear in the latter). This problem has been studied per se in~\cite{GaMaLe21b}, which emphasized the importance in this case of the notion of one-sided tests, namely families of $d$-dependent estimators that asymptotically for large $d$ have a vanishing probability of error under the null hypothesis, and a probability of error bounded away from 1 under the alternative. Conditions for the existence of such tests have been derived in terms of the Kullback-Leibler divergence between the two distributions,
\begin{equation}
    {\rm KL}_d = D(\mathbb P_1^{(d)} || \mathbb P_0^{(d)} \otimes \mathbb P_0^{(d)}) = \sum_{T,T'} \mathbb P_1^{(d)}[T,T'] \ln \left(\frac{\mathbb P_1^{(d)}[T,T']}{\mathbb P_0^{(d)}[T]\mathbb P_0^{(d)}[T']} \right) = \sum_{T,T'} \mathbb P_1^{(d)}[T,T'] \ln L^{(d)}(T,T') \ .
\label{eq_KLd}
\end{equation}
It was indeed shown in~\cite{GaMaLe21b} that ${\rm KL}_d$  is a non-decreasing sequence in $d$, that diverges to $+\infty$ when $d\to +\infty$ if and only if one-sided tests do exist for the corresponding values of $(\lambda,s)$. This leads to the definition of a threshold $s_{\rm c}(\lambda)$ for the tree problem, such that ${\rm KL}_d$ diverges with $d$ if and only if $s>s_{\rm c}(\lambda)$ (we assume again that this property is monotonous in $s$). Some upper and lower bounds on $s_{\rm c}(\lambda)$ have also been proven in~\cite{GaMaLe21b}. The lower bound $s_{\rm c}(\lambda) \ge s_{\rm lb}(\lambda) = 1 / \lambda$, which parallels the one on $s_{\rm algo}$ discussed before, is valid for all $\lambda \ge 1$; for $\lambda \gtrsim 5.5$ the theorem 5 in~\cite{GaMaLe21b} provides an improved lower bound that behaves asymptotically as 
$1 /\sqrt{\lambda}$ for $\lambda \to \infty$. Upper bounds on $s_{\rm c}(\lambda)$ are also provided by theorems 3 and 4 in \cite{GaMaLe21b}, by showing that ${\rm KL}_d$ diverges with $d$ for some values of $(\lambda,s)$; explicit formulas for these bounds are cumbersome to write, see Fig.~\ref{fig:tree_pd} for a graphical representation of one of them. For $\lambda \in [1,1.178]$ the upper and lower bounds coincide, implying that $s_{\rm c}(\lambda)=1/\lambda$ in this interval.

\begin{figure}
    \centering
    \includegraphics[width=0.475\columnwidth]{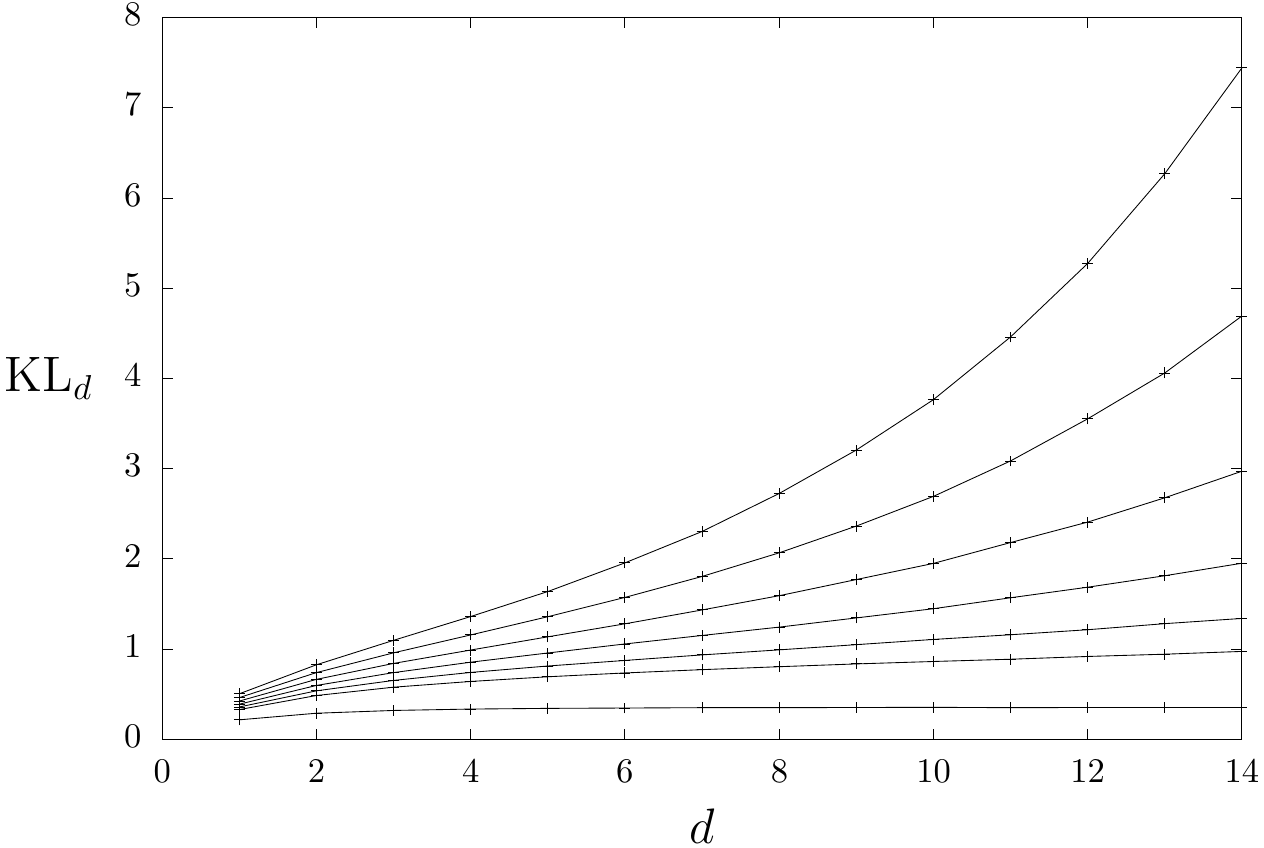}
    \hfill
    \includegraphics[width=0.475\columnwidth]{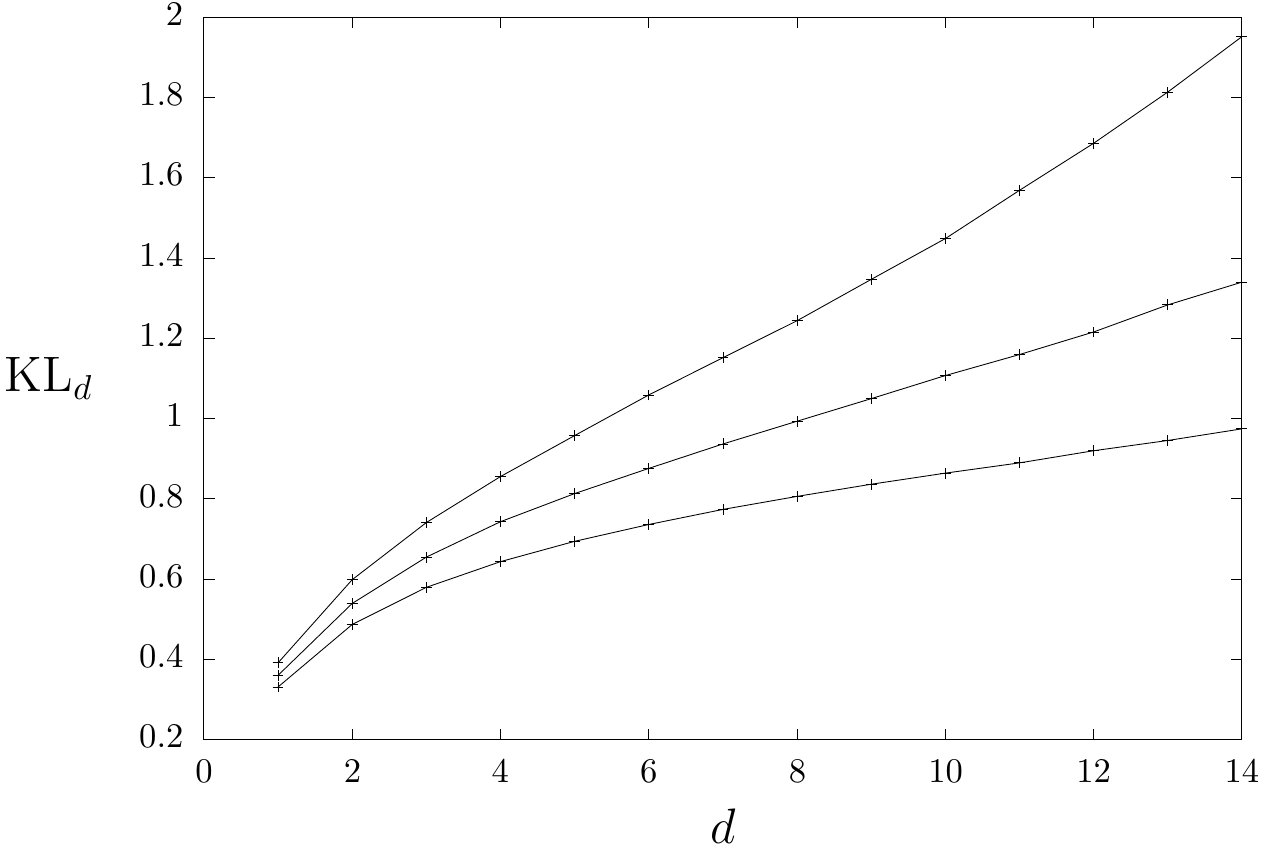}
    \caption{Left panel: numerical estimates of the Kullback-Leibler divergence ${\rm KL}_d$ defined in Eq.~(\ref{eq_KLd}) for the hypothesis testing problem on trees,  as a function of $d$ for $\lambda=1.5$, and from bottom to top $s=0.6$, $s=0.7$, $s=0.72$, $s=0.74$, $s=0.76$, $s=0.78$, and $s=0.8$. Averages are taken over $10^5$ to $10^6$ samples depending on the values of $d$, error bars are of the order of the symbol size. The right panel shows the curves corresponding to $s=0.7$, $s=0.72$ and $s=0.74$ on a different range to better appreciate the transition region.}
    \label{fig:tree_E1logLofd_l1.5}
\end{figure}

We have performed a numerical study of this hypothesis testing problem on trees, through the computation of the Kullback-Leibler divergence ${\rm KL}_d$ defined in Eq.~(\ref{eq_KLd}). We estimated this quantity with an elementary, brute-force procedure, by generating a large number of pairs of trees $(T,T')$ with the law $\mathbb P_1^{(d)}$, computing $L^{(d)}(T,T')$ using the recursions (\ref{eq_Ld}) from the leaves towards the root, and performing the empirical average over the samples. The generation step was done by first drawing the multi-type Galton-Watson tree $\mathcal T$ described in Section \ref{sec_aligned} and then projecting it to the blue and red connected components of the root, $(T,T')=(b({\mathcal T}),r({\mathcal T}))$. The numerical results obtained in this way are presented in Fig.~\ref{fig:tree_E1logLofd_l1.5} for $\lambda=1.5$. One obtains as expected that ${\rm KL}_d$ is a non-decreasing function of $d$, with a behavior at large $d$ that suggests a saturation for small $s$ and a divergence for large $s$. As the range of $d$ that can be investigated is rather small (the limiting factor is the storage of the pair of trees, the memory cost is $O(\lambda^{2d})$ because of the number of vertices that grows exponentially with $d$, with a rate that increases with $\lambda$) one cannot locate affirmatively a transition in a very precise way. Nevertheless it is tempting to conjecture that $s_{\rm c}(\lambda=1.5)\in[0.7,0.74]$, as the curves bend upwards (resp. downwards) for larger (resp. smaller) values of $s$ (see in particular the right panel of Fig.~\ref{fig:tree_E1logLofd_l1.5}), and the divergence is argued in~\cite{GaMaLe21b} to be exponential in $d$ above the transition (we mention as a side remark the open problem of the continuity of this transition, i.e. whether $\lim_{s \to s_{\rm c}^-} \lim_{d \to \infty} {\rm KL}_d$ is finite or not). This procedure to determine $s_{\rm c}$ is somehow subjective and cannot yield very accurate estimates of $s_{\rm c}$, but we did not found a better way to perform the large $d$ extrapolation. We repeated the same analysis for a few different values of $\lambda$; for larger values of $\lambda$ we could only reach $d=10$, whereas for $\lambda=1.2$ we computed ${\rm KL}_d$ up to $d=20$. The results are summarized as a phase diagram in the $(\lambda,s)$ plane in Figure~\ref{fig:tree_pd}, along with the lower bound $\lambda s=1$ and the upper bound from Theorem 3 in~\cite{GaMaLe21b}. 

Let us now come back to the original graph alignment problem and its connection with the tree problem. We first underline the fact that the numerically determined values of $s_{\rm c}(\lambda)$ plotted in Fig.~\ref{fig:tree_pd} are compatible with the estimate $s_{\rm algo}(\lambda) \approx 0.6$ obtained from the finite $n$ simulations of the previous Section in this range of $\lambda$, and we believe that despite the problematic large $d$ extrapolation the determination of $s_{\rm c}(\lambda)$ is more accurate than the one of $s_{\rm algo}(\lambda)$. Furthermore we expect that these two thresholds coincide, namely that $s_{\rm c}(\lambda) = s_{\rm algo}(\lambda)$. As a matter of fact it has been proven in~\cite{GaMaLe21b} that if one-sided tests for the tree problem exist for a given value of $(\lambda,s)$, then a polynomial-time algorithm achieves partial recovery for the corresponding graph problem; however, this is proven for a different algorithm than the one presented above, less efficient but easier to analyze. The intuitive connection between the graph and the tree problem goes as follows. For a given vertex $i$ the algorithm recovers successfully its matched vertex $i'=\bpis(i)$ with a positive probability if the score $L_{i,i'}^{(d)}$ is a maximizer of $L_{i,\cdot}^{(d)}$, and if the degeneracy of this maximum is finite. Forgetting the cycles in the graph and some correlations between overlapping neighborhoods one can picture this vector of $n$ random variables as containing one sample of $L^{(d)}$ under the law $\mathbb{P}_1^{(d)}$, and $n-1$ samples drawn with the law $\mathbb{P}_0^{(d)}$. For the single random variable corresponding to the aligned pair to be larger than the extreme value of the $n-1$ other ones in the large $n$ limit the laws of the likelihood ratio under the null and the alternative have to strongly differ one from the other, with $L$ being typically much larger under $\mathbb{P}_1$ than under $\mathbb{P}_0$, as expressed by the divergence of ${\rm KL}_d$. 

The connection between graph and tree problems is of course a recurrent theme that appeared in the previous literature under many guises. Without attempting to be exhaustive let us give a few examples that will be useful to discuss the situation of the present case. Under the name of the objective method it was shown in~\cite{Al01} that the minimal cost of the matching of a weighted complete graph converges in the large size limit to a quantity that can be computed from an infinite tree. The cavity method~\cite{MePa01} allows to study statistical mechanics models defined on random graphs by exploiting their local convergence to trees. One way to interpret this method is to consider the factor graph associated to the interactions of the original graph model, and to study the latter via the so-called Belief Propagation (BP) algorithm, a message-passing procedure to compute approximations of the local marginals and of the global partition function (via the Bethe free-energy formula), that would be exact if the model was defined on a tree, and that is asymptotically exact for models on graphs locally converging to trees, provided some correlation decay conditions are fulfilled (for simplicity we only discuss here the Replica Symmetric version of the cavity method, see~\cite{MePa01} for a discussion of Replica Symmetry Breaking when this decorrelation condition is violated). In these two examples the limiting object is characterized by a random variable $\boldsymbol X$ (the message passed between adjacent nodes), that obeys a fixed point equation of the form ${\boldsymbol X} \overset{\rm d}{=} g({\boldsymbol X}_1 , \dots ,{\boldsymbol X}_n)$, where the equality is in distribution, and the ${\boldsymbol X}_i$'s are i.i.d. copies of $\boldsymbol X$. This type of fixed point condition is known as a Recursive Distributional Equation (RDE), which can be easily solved numerically by a so-called population dynamics procedure. In the context of inference problems let us also mention the case of the Stochastic Block Model (SBM), where one has to recover a hidden signal made of labels placed on the vertices of a graph, the observations being the edges of the graph, whose probability of presence depends on the labels of the two vertices at its endpoints~\cite{DeKrMoZd11,Mo17,Ab18}. In the sparse regime of the SBM, with constant average degree, the posterior distribution of the labels given the observed graph can be written as a factor graph which converges locally to a tree (treating the information from the absent edges in an average way), and which can be studied with BP and the cavity method. This connects the possibility of efficient recovery of the labels on the graph model to the possibility of (robust) reconstruction on the associated tree problem~\cite{MoNeSl16}. Moreover the limit for the information theoretical possibility of recovery (without consideration of computational efficiency) is deduced from the mutual information between the labels and the observed edges, which itself is expressed in terms of the solution of the RDE via the Bethe free-energy formula. 

We would like to emphasize that the connection between the graph and the tree problem encountered in the present paper exhibits important differences with the well-known examples we have just recalled:
\begin{itemize}
    \item The message passing algorithm has not been obtained through the BP approximation of the factor graph encoding the full posterior of the problem. We replaced instead the posterior by a truncated probability law, that depends on the pair of vertices $i,i'$ for which we estimate the probability that $\bpis(i)=i'$. Once this truncation has been performed the rest of the derivation is asymptotically exact when $d$ is finite with $n \to \infty$. The consequences of this observation are on the one hand that one cannot invoke some correlation decay property to justify the result, and on the other hand that one cannot use the limiting tree problem to compute some thermodynamic quantities like the normalization of the posterior, and deduce from it the information theoretical limits of partial recovery.  

    \item The limiting tree problem is actually a problem involving pairs of trees, not a single tree like for instance in the cavity treatment of the SBM problem. As a consequence the message-passing equations of Eq.~(\ref{eq_MP}) involve messages passed between two pairs of vertices, and not along the edges of a graph as in the usual BP algorithms.

    \item The computations of the observables in the limiting tree problem, notably the ${\rm KL}_d$ of Eq.~(\ref{eq_KLd}), cannot be computed with the usual population dynamics algorithm. As a matter of fact in the recursion of (\ref{eq_Ld}) the array $L_{i,i'} = L^{(d-1)}(T_i,T'_{i'})$ is strongly correlated (all the elements of the $i$-th row share the same tree $T_i$), hence one cannot write a RDE of the form ${\mathbf L}^{(d)} \overset{\rm d}{=} f(\{{\mathbf L}^{(d-1)}_{i,i'} \})$ with i.i.d. copies of a single random variable ${\mathbf L}^{(d-1)}$ in the right hand side. In particular the limit $d \to \infty$ cannot be described explicitly as the solution of a simple fixed-point distributional equation.

\end{itemize}

\section{Conclusions}
\label{sec:conclusions}

Let us conclude by summarizing our main findings and proposing some possible perspectives for further research. It would be desirable to obtain a more accurate determination of the phase transition giving the limit of the successful partial recovery of the hidden permutation by the message-passing algorithm, and to extend it to larger values of $\lambda$. This improvement could rely on better numerical procedures to perform the extrapolation at large $n$ of the simulations on the graph, or at large $d$ for the tree problem numerical experiments. From a more analytical point of view one could hope to either improve the bounds of~\cite{GaMaLe21b} on $s_{\rm c}(\lambda)$ to make them tighter, or to look for simplifications in some limits of the parameters. In the latter perspective, the rather modest dependency of $s_{\rm c}$ and $s_{\rm algo}$ on $\lambda$ in the investigated range, and the shape of the curve in Fig.~\ref{fig:tree_pd}, could lead to the conjecture that $s_{\rm c}(\lambda)$ reaches a strictly positive value in the limit $\lambda \to \infty$. Let us further mention what is at the moment an intriguing numerical coincidence; the authors of~\cite{MaWuXuYu21} studied the detection problem associated to graph alignment, namely the hypothesis testing question of, given a pair of graphs, distinguishing their generation probability between the correlated Erd\H os-R\'enyi law and the product of two independent Erd\H os-R\'enyi laws with the same marginals as the correlated one. They defined an estimator based on the correlation of the number of trees embedded as subgraphs in the observed pairs of graphs, and characterized the range of parameters for which this estimator achieves asymptotically a vanishing probability of error under both hypotheses; in the constant-degree regime this happens for all $\lambda > 0$ whenever $s > \sqrt{\alpha} $, where $\alpha$ is the Otter's constant~\cite{Ot48} that governs the rate of growth of the number of unlabelled trees with the number of vertices (the results of~\cite{MaWuXuYu21} actually cover also denser regimes with degrees diverging with $n$). The numerical value of this threshold is $\sqrt{\alpha}\approx 0.5817$, indicated with an horizontal dot-dashed line in Fig.~\ref{fig:tree_pd}, just slightly below the typical values of $s_{\rm c}$ we observed,  which makes this constant $\sqrt{\alpha}$ a possible candidate for the conjectured limit of 
$s_{\rm c}(\lambda)$ when $\lambda \to \infty$.

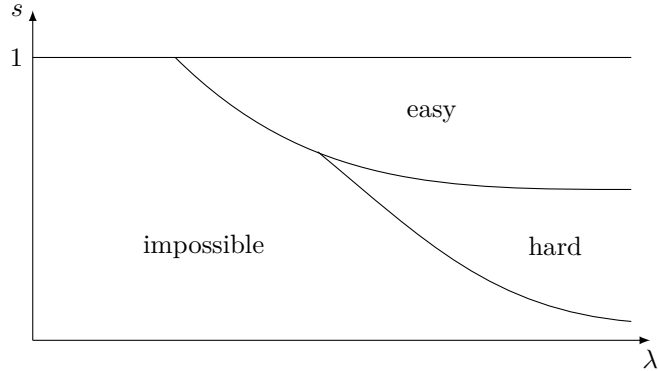
\begin{figure}
    \begin{center}
    \begin{tikzpicture}[scale=1.25]
    \draw[-latex] (0,0) -- (0,3.5) node [left] {$s$};
    \draw[-latex] (0,0) -- (6.5,0) node [below] {$\lambda$};
    \draw (6.3,3) -- (0,3) node [left] {$1$};
    \draw (1.5,3) to[out=-45,in=180] (6.3,1.6);
    \draw (3,2.0) to[out=-40,in=175] (6.3,.2);
    \draw (1.8,1) node {impossible};
    \draw (4.2,2.4) node {easy};
    \draw (5.5,1) node {hard};
    
    \end{tikzpicture}
    \end{center}
    \caption{A sketch of the conjectured phase diagram for the alignment problem of correlated Erd\H os-R\'enyi random graphs.}
    \label{fig:pd_conjectured}
\end{figure}

The sketch of Fig.~\ref{fig:pd_conjectured} summarizes the conjectured phase diagram for the partial recovery of the hidden permutation in the constant degree regime of the correlated Erd\H os-R\'enyi ensemble. In the easy phase partial recovery is achievable by a polynomial-time algorithm, in the hard phase it is achievable but in an a priori exponential time, while in the impossible phase the information contained in the graphs is insufficient to recover a constant fraction of the hidden permutation, even without bounds on the computational power employed. The evidences in favor of this shape of the phase diagram are on the one hand the numerical results presented in this paper, the boundary of the easy phase corresponding to the threshold $s_{\rm algo}(\lambda)$, and on the other hand the various bounds previously obtained in the literature:  $\lambda s < 1 $ is a sufficient condition to be in the impossible phase~\cite{GaMaLe21a}, for $\lambda s > 4 $ partial recovery is information-theoretically possible~\cite{WuXuYu21}, hence this regime corresponds to an easy or hard phase. As the lower bound on $s_{\rm c}$ given by theorem 5 in~\cite{GaMaLe21b} crosses the line $\lambda s = 4$ for $\lambda$ large enough, a hard phase must appear in this regime. On the contrary at small $\lambda$, or more precisely for $\lambda \in [1,1.178]$ the bounds of~\cite{GaMaLe21b} are tight, hence the initial portion of the line $\lambda s = 1$ corresponds to a transition between the impossible and the easy phase. The upper and lower bounds on the information theoretical limit of partial recovery are relatively far apart one from the other, which leaves some room for the location of the transition line between the hard and impossible phases, and of the "triple point" where the three phases meet. One can in particular wonder about the interval of $\lambda$ for which the curve $\lambda s=1$ is the boundary of the impossible phase.

The message-passing algorithm studied in this paper was obtained by truncating the posterior, discarding the information outside the neighborhoods of radius $d$ around the considered vertices. One could have hoped that when $d$ is large enough, in particular for $d=\Theta(\ln n)$, the information discarded becomes negligible, as the estimator $\pih$ is then computed from the observation of a finite fraction of the graphs. The existence of a hard phase for this algorithm shows that this expectation is wrong, some global information that is present in the full posterior distribution is lost in the local computation, for any arbitrary large radius of observation. Note that the notion of a hard phase is a priori related to a specific algorithm; we expect, however, that the one investigated in this paper is optimal among all local procedures, and maybe more generically for all polynomial-time algorithms for partial recovery in the sparse regime.

A possible direction for future work would be to consider ensembles of correlated random graphs more generic than the Erd\H os-R\'enyi one. In \ref{app:genensembles} we introduce such an ensemble that allows to tune in a flexible way the degree distribution of the generated graphs, as well as the level of correlation between them, and show that the message-passing algorithm can be generalized to this case. A particularly challenging situation is the one of regular graphs: it is known from~\cite{KiSuVu02} that random regular graphs do not have non-trivial automorphisms, with high probability. Hence perfect recovery is asymptotically possible for a pair of regular graphs in the noiseless case (i.e. when one observes a regular graph and a reshuffled version of it), as among all the permutations only the hidden one will achieve a perfect alignment of the two graphs; this is of course only an information-theoretic statement, the exhaustive search among all the permutations being computationally inefficient. This opens several questions that, to the best of our knowledge, have not been treated previously: can partial recovery be achieved for some pairs of partially correlated regular graphs? Can these tasks (perfect or partial recovery for noiseless or noisy regular graphs) be performed in a computationally efficient way? Note that the generalized message-passing algorithm presented in the \ref{app:genensembles} is completely useless on regular graphs, all trees of non-backtracking walks being regular at any depth there is strictly no local information to be exploited in this way.

Among other ensembles of correlated pairs of graphs let us also mention the case of the correlated Stochastic Block Model, whose perfect recovery in the logarithmic degree regime has been studied in~\cite{OnGaEr16,RaSr21}. One could also investigate these problems in the constant degree regime, and in particular study the interplay between the partial recovery of the hidden permutation and of the hidden labels. Finally there exists several "seeded" versions of the alignment problem~\cite{YaGr13,LyFiPr14,MoXu20}, that are often relevant in applications, in which a side information on the hidden permutation is provided to the observer in addition to the pair of graphs, with either a part of the permutation being revealed, or under the form of an affinity matrix that favors the matching of some pairs of vertices, or some indications on strictly forbidden matchings between some subset of vertices. It should be possible to adapt the message-passing algorithm to exploit this additional information, by incorporating the latter in the truncated posterior.

\section*{Acknowledgments}

We thank Luca Ganassali and Marc Lelarge for useful discussions. 

\appendix
\addtocontents{toc}{\fixappendix}

\section{Generalized ensembles}\label{app:genensembles}

Correlated Erd\H os-R\'enyi random graphs have Poissonian degree distributions; however, in practical applications one often encounters the problem of aligning graphs with degree distributions significantly distinct from Poissonian. It is thus desirable to have a model of correlated random graphs with some flexibility in the degree distribution. One possibility is to draw a parent graph with an arbitrary degree distribution, and then subsample its edge set twice independently (see for instance~\cite{YuXuLi21} for an example of such a construction with power-law degree distributions). In this Appendix we propose a model that allows for a finer control of the degree distribution of the correlated pair of graphs, and show that the message passing algorithm described in the main text for the ER case can be adapted to this generalized ensemble. We shall follow essentially the same steps as in Sections \ref{sec:corrgraphs} and \ref{sec:inference}, namely define the correlated graph model, study its local behavior, and deduce from it an algorithmic procedure for its alignment.

\subsection{Random graphs with prescribed degree distributions}

For the sake of clarity, let us start by recalling some well-known results on a procedure that allows to draw random graphs with a prescribed degree distribution, called the \textit{configuration model}. Suppose that a distribution $P(l)$, admitting a finite second moment, is given on the non-negative integers, and we are asked to generate a random graph on $n$ vertices with an empirical degree distribution that is close to $P$ in the large size limit. A simple way to achieve this goal is to draw $n$ degrees $l_1,\dots,l_n$ independently from the law $P$, and associate to each vertex $i$ a number $l_i$ of half-edges. Subsequently, we draw a uniformly random pairing of the half-edges to build the edges of the random graph. Some pairings will produce self-loops and multiple edges between pairs of vertices, but it is possible to show that the probability of generating a simple graph (i.e., a graph without self-loops and multiple edges) with this procedure remains positive in the large $n$ limit, hence a finite (on average) number of rejections will eventually lead to a simple graph with the correct degree distribution.

The local properties of such random graphs bear some similarities with the ones of the ER model: the depth-$d$ neighborhood of an arbitrarily chosen vertex converges with high probability to a random tree $T$ when $n \to \infty$ with $d$ finite. To describe the law of $T$ let us first define another distribution on the integers,
\begin{equation}
    \widehat{P}(l) = \frac{(l+1)P(l+1)}{\sum_{l'=0}^\infty l'P(l') } \ ,
\end{equation}
which corresponds to the size-biased version of $P$, also called the edge perspective degree distribution in this context. We can now characterize the law of the random tree $T$ as follows: its root has a number $l$ of offsprings with probability $P(l)$, each of these offsprings being the root of an independent copy of a random tree $\widehat{T}$, where in $\widehat{T}$ all vertices (including the root) have offspring distribution $\widehat{P}$. The fact that the root of $T$ has degree distribution $P$ follows directly from the definition of the ensemble. On the other hand, all other vertices are reached by crossing an edge, and in this exploration process the probability to end up in a vertex of degree $l+1$ is proportional to the number of half-edges around such vertices, which is itself proportional to $(l+1)P(l+1)$.

\subsection{Correlated random graphs with prescribed degree distributions}

Let us now introduce a model for the generation of a pair $(\bA,\bC)$ of correlated random graphs with some prescribed degree distributions. The generation process depends on the choice of a joint distribution for three non-negative integers, $q(\lb, \lr, \lbi)$, which we assume has finite second moments, and fulfills the property $q(\lb, \lr, \lbi)=q(\lr, \lb, \lbi)$. For each vertex $i \in [n]$ one draws a triplet $(l_{{\rm b},i},l_{{\rm r},i},l_{{\rm bi},i} )$ i.i.d. from the law $q$, in such a way that, from vertex $i$, a number $l_{{\rm b},i}$ (resp. $l_{{\rm r},i}$, $l_{{\rm bi},i}$) of blue (resp. red, bicolored) half-edges emerges. One then draws three uniform pairings of these three type of half-edges, in order to produce a colored graph $\bG$ with three types of edges (if $\bG$ is not simple, the generation process is restarted). A pair of graphs $(\bA,\bC)$ is finally produced from $\bG$ by keeping in $\bA$ (resp. in $\bC$) the blue and bicolored (resp. red and bicolored) edges of $\bG$. The degree distribution of $\bA$ and $\bC$ is then easily seen to be
\begin{equation}
    P(l) \coloneqq \sum_{\lbi=0}^l \sum_{\lr=0}^\infty q(l-\lbi,\lr,\lbi) \ ;
    \label{eq_P_from_q}
\end{equation}
by virtue of the property $q(\lb, \lr, \lbi)=q(\lr, \lb, \lbi)$, the graph $\bA$ and $\bC$ have the same degree distribution. As we shall see below the law of $\bA$ is not in general the same as the single configuration model with degree distribution $P$.

One can then create a graph $\bB$ with $\bC_{ij}=\bB_{\bpis(i),\bpis(j)}$ where $\bpis$ is a uniformly random permutation, and study the inference problem of recovering $\bpis$ from the observation of $(\bA,\bB)$. As in the ER case we will propose an approximate estimator of $\bpis$ based on the local properties of the graphs, that we shall now describe. In order to do this we define the following size-biased versions of $q$:
\begin{align}
    \hq(\lb, \lr, \lbi) &\coloneqq \frac{(\lbi +1) \, q(\lb, \lr, \lbi+1)}{\underset{\lbp, \lrp, \lbip}{\sum} \lbip \, q(\lbp, \lrp, \lbip)} \ , \\
    \tq(\lb, \lr, \lbi) &\coloneqq \frac{(\lb +1) \, q(\lb+1, \lr, \lbi)}{\underset{\lbp, \lrp, \lbip}{\sum} \lbp \, q(\lbp, \lrp, \lbip)}\ , \\
    \dq(\lb, \lr, \lbi) &\coloneqq \tq(\lr, \lb, \lbi) \ .
\end{align}
Let us call $\mathcal T$ the colored branching process obtained by a local exploration of $\bG$ from an arbitrary root vertex. Generalizing the construction given in the main text for the ER case, and recalling the explanations on the appearance of size-biased versions of the degree distribution given above in the single graph case, one realizes that the law of $\mathcal T$ can be described as follows: the root of $\mathcal T$ has $\lb$ offsprings linked to it by a blue edge, $\lr$ by a red edge and $\lbi$ by a bicolored edge, with probability $q(\lb,\lr,\lbi)$. Each of these $\lb$ (resp. $\lr$, $\lbi$) offsprings is the root of an independent copy of the tree $\widetilde{\mathcal T}$ (resp. $\dot{\mathcal T}$, $\widehat{\mathcal T}$), where the law of $\widetilde{\mathcal T}$ (resp. $\dot{\mathcal T}$, $\widehat{\mathcal T}$) follows exactly the same definition as the one of $\mathcal T$ with the replacement of the distribution $q$ by $\tq$ (resp. $\dq$, $\hq$).

Consider now the question investigated in Section \ref{sec:local_disaligned} in the ER case, namely the joint law of $(T,T')$ of the neighborhood $T$ of $i$ in $\bA$ and $T'$ of $i'$ in $\bB$, where $i$ and $i'$ are aligned vertices. As in the main text $(T,T')=(b(\mathcal T),r(\mathcal T))$, where $b(\mathcal T)$ (resp. $r(\mathcal T)$) returns the connected component of the root of the blue and bicolored (resp. red and bicolored) edges of $\mathcal T$, neighborhood of $i$ in $\bG$. From the recursive description of the law of $\mathcal T$ given above we can deduce a recursive expression of the law of $(T,T')$. More precisely, we shall denote $\mathbb{P}_1^{(d)}[T,T']$ the law of $(b({\mathcal T}),r({\mathcal T}))$, $\widehat{\mathbb{P}}_1^{(d)}[T,T']$ the law of $(b(\widehat{{\mathcal T}}),r(\widehat{{\mathcal T}}))$, and $\widetilde{\mathbb{P}}_1^{(d)}[T,T']$ the law of $(b(\widetilde{{\mathcal T}}),r(\widetilde{{\mathcal T}}))$, with $\mathcal T$, $\widehat{\mathcal T}$ and $\widetilde{\mathcal T}$ the random colored trees defined previously, when observed up to the $d$-th generation. We will also denote $\mathbb{P}_0^{(d)}[T]$, $\widehat{\mathbb{P}}_0^{(d)}[T]$ and $\widetilde{\mathbb{P}}_0^{(d)}[T]$ their marginals obtained by summing over $T'$. The translation of the recursive definition of $\mathcal T$, exploiting the fact that once a monochromatic edge of $\mathcal T$ has been crossed one can discard the edges of the opposite color below it, and the symmetry between the colors, yields:
\begin{multline}
\mathbb{P}_1^{(d)}[T,T']  = 
\sum_{l_{\rm b},l_{\rm r},l_{\rm bi}=0}^\infty  q(\lb,\lr,\lbi) \, \mathbb{I}(l=l_{\rm b} + l_{\rm bi}) \ \mathbb{I}(l'=l_{\rm r} + l_{\rm bi}) \\
\sum_{Q_1,\dots,Q_{l_{\rm b}}} \sum_{R_1,\dots,R_{l_{\rm r}}} \sum_{S_1,S'_1,\dots , S_{l_{\rm bi}},S'_{l_{\rm bi}}} \prod_{i=1}^{l_{\rm b}} \widetilde{\mathbb{P}}_0^{(d-1)}[Q_i] \prod_{i=1}^{l_{\rm r}} \widetilde{\mathbb{P}}_0^{(d-1)}[R_i] \prod_{i=1}^{l_{\rm bi}} \widehat{\mathbb{P}}_1^{(d-1)}[S_i,S'_i]  \\ 
\comprimi\frac{1}{l! l'!} \sum_{\pi,\pi'} \mathbb{I}((T_1,\dots,T_l)=\pi(Q_1,\dots,Q_{l_{\rm b}} ,S_1,\dots,S_{l_{\rm bi}})) \, \mathbb{I}((T'_1,\dots,T'_{l'})=\pi'(R_1,\dots,R_{l_{\rm r}},S'_1,\dots,S'_{l_{\rm bi}})) \ , \nonumber
\end{multline}
where $l$ (resp. $l'$) is the degree of the root of $T$ (resp. $T'$), and $T_1,\dots,T_l$ (resp. $T'_1,\dots,T'_{l'}$) are the subtrees rooted below it,
$\pi$ (resp. $\pi'$) is a permutation of its $l$ (resp. $l'$) arguments, and we use the convention $\mathbb{P}_1^{(0)}[T,T']=1$. This can be simplified into  a generalization of the equation (\ref{eq_P1}) of the ER case,
\begin{equation}\comprimi
\mathbb{P}_1^{(d)}[T,T']  = \sum_{l_{\rm bi}=0}^{\min(l,l')}
\frac{q(l-\lbi,l'-\lbi,\lbi)}{\binom{l}{l_{\rm bi}} \binom{l'}{l_{\rm bi}} l_{\rm bi}!}  \sum_{I,I',\sigma} \prod_{i \in I} \widehat{\mathbb{P}}_1^{(d-1)}[T_i,T'_{\sigma(i)}] \prod_{i \in [l] \setminus I} \widetilde{\mathbb{P}}_0^{(d-1)}[T_i] \prod_{i \in [l'] \setminus I'} \widetilde{\mathbb{P}}_0^{(d-1)}[T'_i] \ , \label{eq_P1_generalized}
\end{equation}
where $I$ (resp. $I'$) is a subset of $[l]$ (resp. of $[l']$) of $l_{\rm bi}$ elements, and $\sigma$ a bijection from $I$ to $I'$. The expressions of $\widehat{\mathbb{P}}_1^{(d)}[T,T']$ and $\widetilde{\mathbb{P}}_1^{(d)}[T,T']$ are of the same form, the only modification being the replacement of the law $q$ by $\hq$ and $\tq$ in the right hand side of (\ref{eq_P1_generalized}).

Summing over $T'$ yields a recursive expression of the marginal law,
\begin{equation}
\mathbb{P}_0^{(d)}[T]  =  \sum_{l_{\rm bi}=0}^l
\frac{q(l-\lbi,\lbi)}{\binom{l}{l_{\rm bi}} }  \sum_{I} \prod_{i \in I} \widehat{\mathbb{P}}_0^{(d-1)}[T_i] \prod_{i \in [l] \setminus I} \widetilde{\mathbb{P}}_0^{(d-1)}[T_i]  \ ,
\label{eq_P0_generalized}
\end{equation}
where here and in the following a law $q$, $\hq$ and $\tq$ with only two arguments is understood to be the the marginal when the number of red edges has been discarded, e.g.
\begin{equation}
    q(\lb, \lbi) = \sum_{\lr=0}^\infty q(\lb, \lr, \lbi) \ .
\end{equation}
As in the joint case $\widehat{\mathbb{P}}_0^{(d)}[T]$ and $\widetilde{\mathbb{P}}_0^{(d)}[T]$ are given by (\ref{eq_P0_generalized}) with the law $q$ in the right hand side replaced by $\hq$ and $\tq$ respectively.

For a generic choice of the distribution $q$ the size-biased versions $\hq(\lb,\lbi)$ and $\tq(\lb,\lbi)$ differ, hence $\widehat{\mathbb{P}}_0^{(d)} \neq \widetilde{\mathbb{P}}_0^{(d)}$; in this case (\ref{eq_P0_generalized}) cannot be written as the product of the degree distribution $P(l)$ multiplied by the product of the probabilities of the subtrees, which shows that the law of $\bA$ is not the same as the one of the configuration model with the degree distribution $P(l)$ defined in (\ref{eq_P_from_q}). If instead $q$ is such that $\hq(\lb,\lbi)=\tq(\lb,\lbi)$, then by induction on $d$ one finds that $\widehat{\mathbb{P}}_0^{(d)}= \widetilde{\mathbb{P}}_0^{(d)}$, hence one can simplify (\ref{eq_P0_generalized}) and its counterpart for $\widehat{\mathbb{P}}_0^{(d)}$ into
\begin{equation}
   \mathbb{P}_0^{(d)}[T]  =  \left( \sum_{l_{\rm bi}=0}^l
q(l-\lbi,\lbi) \right) \prod_{i=1}^l  \widehat{\mathbb{P}}_0^{(d-1)}[T_i] \ , \qquad \widehat{\mathbb{P}}_0^{(d)}[T]  =  \left( \sum_{l_{\rm bi}=0}^l
\hq(l-\lbi,\lbi) \right) \prod_{i=1}^l  \widehat{\mathbb{P}}_0^{(d-1)}[T_i] \ ,
\end{equation}
which corresponds to the branching process description of the configuration model with the degree distribution $P(l)$ of \eqref{eq_P_from_q}. A short computation reveals indeed that $\sum_{l_{\rm bi}=0}^l
\hq(l-\lbi,\lbi)$ is the size bias of $\sum_{l_{\rm bi}=0}^l q(l-\lbi,\lbi)$ in this case.

These simplifications occur in particular when $q$ is a product of three Poisson distributions for the three types of degrees: as a Poisson distribution is equal to its size-bias one has then $q=\hq=\tq=\dq$ and the equations (\ref{eq_P0_generalized}) and \eqref{eq_P1_generalized} reduce to their counterparts \eqref{eq_P0} and \eqref{eq_P1} derived in the ER model.

\subsection{Generalization of the message passing algorithm}

Consider now the inference problem of recovering the unknown permutation $\bpis$ from the observation of $(\bA,\bB)$ generated with the correlated graph ensemble for an arbitrary distribution $q$. The message passing algorithm described in the main text for the correlated ER ensemble can be naturally extended to handle this case. As a matter of fact the reasoning linking the truncated posterior probability $\bpis(i)=i'$ to the likelihood ratio of the neighborhoods $(T,T')$ of the vertices $i$ and $i'$ formalized in \eqref{eq_truncated_posterior} and \eqref{eq_ratio} applies verbatim to the generalized ensemble, thanks to its similar local convergence properties towards random trees. A slight complication occurs because in general the null hypothesis $\mathbb{P}_0^{(d)}[T]$ does not factorize as a product over its subtrees, hence one cannot write directly a recursion relation on $L^{(d)}(T,T')$, but in any case both $\mathbb{P}_0^{(d)}[T]$ and $\mathbb{P}_1^{(d)}[T,T']$ can be computed recursively thanks to \eqref{eq_P1_generalized} and \eqref{eq_P0_generalized}). To describe in a more compact way the algorithm let us first define a function $f_1$ that takes as input two integers $l$ and $l'$, an $l \times l'$ array $M$, a $l$-dimensional vector $u$ and a $l'$-dimensional vector $u'$, as
\begin{equation}
f_1(l,l';M,u,u')  = \sum_{l_{\rm bi}=0}^{\min(l,l')}
\frac{q(l-\lbi,l'-\lbi,\lbi)}{\binom{l}{l_{\rm bi}} \binom{l'}{l_{\rm bi}} l_{\rm bi}!}  \sum_{I,I',\sigma} \prod_{i \in I} M_{i,\sigma(i)} \prod_{i \in [l] \setminus I} u_i \prod_{i \in [l'] \setminus I'} u'_i \ ,
\end{equation}
where the meaning of $I$, $I'$ and $\sigma$ are the same as in \eqref{eq_P1_generalized}. The function $\widehat{f}_1(l,l';M,v,v') $ is defined by the same expression where $q$ is replaced by $\hq$. We further define a function $f_0$ with arguments an integer $l$ and two $l$-dimensional vectors $u$ and $v$, according to
\begin{equation}
f_0(l;u,v)  = \sum_{l_{\rm bi}=0}^l
\frac{q(l-\lbi,\lbi)}{\binom{l}{l_{\rm bi}} }  \sum_{I} \prod_{i \in I} u_i \prod_{i \in [l] \setminus I} v_i  \ ,
\end{equation}
along with $\widehat{f}_0$ and $\widetilde{f}_0$ defined similarly with $q$ replaced by $\hq$ and $\tq$ respectively.

The generalized algorithm uses the following sets of messages:
\begin{itemize}
    \item $m_{ii' \to jj'}^{(t)}$ for all vertices $i$ of $A$, all vertices $i'$ of $B$, all $j \in \partial i$, and all $j' \in \partial i'$;
    \item $\widehat{m}_{i \to j}^{(t)}$ and $\widetilde{m}_{i \to j}^{(t)}$ for all vertices $i$ of $A$ and all $j \in \partial i$;
    \item $\widehat{m}_{i' \to j'}^{(t)}$ and $\widetilde{m}_{i' \to j'}^{(t)}$ for all vertices $i'$ of $B$ and all $j' \in \partial i'$.
\end{itemize}
All of them are initialized to $1$ at $t=0$, and computed by induction for $t=1,\dots,d-1$ as:
\begin{align}
m_{ii' \to jj'}^{(t)} &\comprimi = \widehat{f}_1(d_i-1,d_{i'}-1;\{m_{kk' \to ii'}^{(t-1)} \}, \{\widetilde{m}_{k \to i}^{(t-1)} \} , 
\{\widetilde{m}_{k' \to i'}^{(t-1)} \} \colon k \in \partial i \setminus j \, , \,  k' \in \partial i' \setminus j' \} ) \ , \\
\widehat{m}_{i \to j}^{(t)} & = \widehat{f}_0(d_i-1;\{\widehat{m}_{k \to i}^{(t-1)} \} , \{\widetilde{m}_{k \to i}^{(t-1)} \} \colon k \in \partial i \setminus j) \ , \\
\widetilde{m}_{i \to j}^{(t)} & = \widetilde{f}_0(d_i-1;\{\widehat{m}_{k \to i}^{(t-1)} \} , \{\widetilde{m}_{k \to i}^{(t-1)} \} \colon k \in \partial i \setminus j) \ , \\
\widehat{m}_{i' \to j'}^{(t)} & = \widehat{f}_0(d_{i'}-1;\{\widehat{m}_{k' \to i'}^{(t-1)} \} , \{\widetilde{m}_{k' \to i'}^{(t-1)} \} \colon k' \in \partial i' \setminus j') \ , \\
\widetilde{m}_{i' \to j'}^{(t)} & = \widetilde{f}_0(d_{i'}-1;\{\widehat{m}_{k' \to i'}^{(t-1)} \} , \{\widetilde{m}_{k' \to i'}^{(t-1)} \} \colon k' \in \partial i' \setminus j') \ .
\end{align}
All messages are assumed to be initialized to $1$. Finally the scores are given by
\begin{align}\label{app:score}
L_{ii'}^{(d)} &= \frac{f_1(d_i,d_{i'};\{m_{jj' \to ii'}^{(d-1)} \}, \{\widetilde{m}_{j \to i}^{(d-1)} \} , 
\{\widetilde{m}_{j' \to i'}^{(d-1)} \} \colon j \in \partial i \, , \,  j' \in \partial i' \} )}{f_0(d_i;\{\widehat{m}_{j \to i}^{(d-1)} \} , \{\widetilde{m}_{j \to i}^{(d-1)} \} \colon j \in \partial i) \ f_0(d_{i'};\{\widehat{m}_{j' \to i'}^{(d-1)} \} , \{\widetilde{m}_{j' \to i'}^{(d-1)} \} \colon j' \in \partial i')  }\ ,
\end{align}
from which the estimator $\pih$ is constructed by matching each vertex $i$ with the vertex $i'$ maximizing $L_{ii'}^{(d)}$, which implements exactly the same strategy than in the ER case.

Note that the flexibility of the random graph ensemble presented in this Appendix allows in particular to generate pairs of $k$-regular random graphs with a law that interpolates smoothly between the completely uncorrelated case (when $q(\lb,\lr,\lbi)=\delta_{\lb,k} \delta_{\lr,k} \delta_{\lbi,0}$) and the perfectly correlated case where the two graphs are identical (with $q(\lb,\lr,\lbi)=\delta_{\lb,0} \delta_{\lr,0} \delta_{\lbi,k}$). However, in this regular case the message passing algorithm is completely inefficient: all trees of non-backtracking walk are regular for any value of $d$ and around any vertex, the scores between any pair of vertices will thus be all equal, hence contain no information on the signal $\bpis$.

\subsection{Weighted random graphs}

The matrix alignment problem, studied for instance in~\cite{Ding2018,GaLeMa19,Fan2019,WuXuYu21,Ga20}, concerns the recovery of an unknown permutation $\bpis$ from the observation of a pair of matrices $(\bA,\bB)$, where $\bB$ is obtained from $\bC$ through the action of $\bpis$ on the row and column indices, while the matrix elements $(\bA_{ij},\bC_{ij})$ are pairs of correlated random variables. It corresponds thus to the graph alignment problem for weighted complete graphs. 

We shall now briefly discuss a further generalization of our approach to the case of sparse correlated weighted graphs. Let us consider a joint probability density $\rho(w,w')$, symmetric under the exchange of its arguments, and with  a marginal law $\rho_{\rm m}(w)=\int \dd w' \rho(w,w')$. We modify the correlated random graph ensembles as follows. From the colored random graph $\bG$ (drawn either from the ER ensemble or from its generalization with a prescribed degree distribution) we derive a pair of weighted graphs $(\bA,\bC)$ by, for each pair $i<j$ of vertices:
\begin{itemize}
    \item drawing $(\bA_{ij},\bC_{ij})$ from $\rho$ if the edge $\{i,j\}$ is bicolored in $\bG$;
    \item drawing $\bA_{ij}$ from $\rho_{\rm m}$ and setting $\bC_{ij}=0$ if the edge $\{i,j\}$ is blue in $\bG$;
    \item drawing $\bC_{ij}$ from $\rho_{\rm m}$ and setting $\bA_{ij}=0$ if the edge $\{i,j\}$ is red in $\bG$;
    \item setting $\bA_{ij}=\bC_{ij}=0$ if the edge $\{i,j\}$ is absent from $\bG$.
\end{itemize}
The description of the local properties of the graphs $(\bA,\bB)$ can then be adapted to this weighted setting. In particular the law $\mathbb{P}_1^{(d)}[T,T']$ for the (now weighted) tree neighborhoods of two aligned vertices in $\bA$ and $\bB$ admit a recursive decomposition that follows from the inclusion of the weight distribution in \eqref{eq_P1_generalized}, namely
\begin{multline}
\mathbb{P}_1^{(d)}[T,T']  = \sum_{l_{\rm bi}=0}^{\min(l,l')}
\frac{q(l-\lbi,l'-\lbi,\lbi)}{\binom{l}{l_{\rm bi}} \binom{l'}{l_{\rm bi}} l_{\rm bi}!}  
\\ 
\sum_{I,I',\sigma} \prod_{i \in I} \rho(w_i,w'_{\sigma(i)}) \widehat{\mathbb{P}}_1^{(d-1)}[T_i,T'_{\sigma(i)}] \prod_{i \in [l] \setminus I} \rho_{\rm m}(w_i) \widetilde{\mathbb{P}}_0^{(d-1)}[T_i] \prod_{i \in [l'] \setminus I'} \rho_{\rm m}(w'_i) \widetilde{\mathbb{P}}_0^{(d-1)}[T'_i] \ , 
\end{multline}
where $w_i$ (resp., $w'_i$) is the weight of the edge between the root of $T$ (resp., of $T'$) and its $i$-th offspring. The message-passing algorithm can then be straightforwardly adapted to incorporate the information coming from these weights. Denoting $w_{ij}$ and $w'_{i'j'}$ the weights on the edges of the observed graphs $A$ and $B$, the following update equations can be derived
\begin{equation}
\resizebox{0.98\columnwidth}{!}{$
\begin{split}
m_{ii' \to jj'}^{(t)} &
        =\comprimi \rho(w_{ij}\!,\!w_{i'j'}')\widehat{f}_1(d_i-1,d_{i'}-1;\{m_{kk' \to ii'}^{(t-1)} \}, \{ \widetilde{m}_{k \to i}^{(t-1)} \}\ , 
\{\widetilde{m}_{k' \to i'}^{(t-1)} \} \colon k \in \partial i\!\!\setminus\!\! j \, , \,  k' \in \partial i'\!\!\setminus\!\! j' ), \\
\widehat{m}_{i \to j}^{(t)} & = \rho_{\rm m}(w_{ij})\widehat{f}_0(d_i-1;\{\widehat{m}_{k \to i}^{(t-1)} \} , \{\widetilde{m}_{k \to i}^{(t-1)} \} \colon k \in \partial i \setminus j) \ , \\
\widetilde{m}_{i \to j}^{(t)} & = \rho_{\rm m}(w_{ij})\widetilde{f}_0(d_i-1;\{\widehat{m}_{k \to i}^{(t-1)} \} , \{\widetilde{m}_{k \to i}^{(t-1)} \} \colon k \in \partial i \setminus j) \ , \\
\widehat{m}_{i' \to j'}^{(t)} & = \rho_{\rm m}(w_{i'j'}')\widehat{f}_0(d_{i'}-1;\{\widehat{m}_{k' \to i'}^{(t-1)} \} , \{ \widetilde{m}_{k' \to i'}^{(t-1)} \} \colon k' \in \partial i' \setminus j') \ , \\
\widetilde{m}_{i' \to j'}^{(t)} & = \rho_{\rm m}(w_{i'j'}')\widetilde{f}_0(d_{i'}-1;\{  \widehat{m}_{k' \to i'}^{(t-1)} \} , \{ \widetilde{m}_{k' \to i'}^{(t-1)} \} \colon k' \in \partial i' \setminus j') \ ,
\end{split}
$}
\end{equation}
with initial conditions $\widehat{m}^{(0)}_{i\to j}=\widetilde{m}^{(0)}_{i\to j}=\rho_{\rm m}(w_{ij})$, $\widehat{m}^{(0)}_{i'\to j'}=\widetilde{m}^{(0)}_{i'\to j'}=\rho_{\rm m}(w'_{i'j'})$, $m_{ii' \to jj'}^{(0)}=\rho(w_{ij},w_{i'j'}')$, while the computation of the scores keeps the form in \eqref{app:score}, from which the estimator can be built by the row maximization procedure.

Note that when $\rho(w,w')=\rho_{\rm m}(w) \rho_{\rm m}(w')$ the contributions of the weights cancel out in the computation of the score, which coincides then with the unweighted computation. Indeed in this case the weights bring no information on whether the edges were aligned or not. The case $\rho(w,w')=\rho_{\rm m}(w)\delta(w-w')$ with $\rho_{\rm m}$ absolutely continuous is the opposite limit: edges with equal weights in $A$ and $B$ were certainly bicolored in $G$ and aligned, those with different weights are certainly not (this would signal itself as zeros and formal infinities in the message passing algorithm). The level of correlation in $\rho$ allows to tune the amount of information on the alignment provided by the weights between these two limit cases.

\section{Further numerical results}
\label{app:numerical}

In this Appendix we present a series of additional results obtained from the numerical simulations of the message-passing algorithm on correlated Erd\H os-R\'enyi random graphs.

\subsection{A comparison of the scores between pairs of aligned and quasi-aligned vertices}
\label{app:scores}

\begin{figure}
    \centering
    \includegraphics{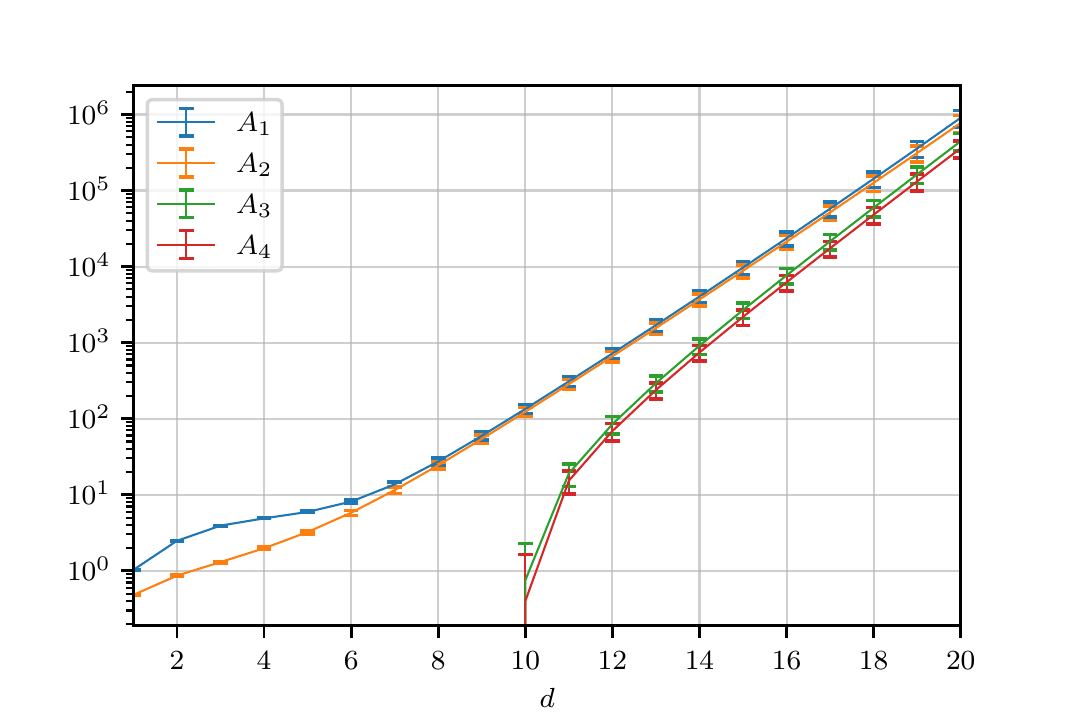}
    \caption{The various averages of logarithm of scores $A_1$, $A_2$, $A_3$, $A_4$ defined in Eqs.~(\ref{eq:argmax_score}-\ref{eq:rand_score}) as a function of the depth $d$, for system size $n=2048$, average degree $\lambda=2.9$, correlation $s=0.79$. Each point is averaged over $15$ independent realizations of the two graphs $\bA, \bB$.} 
    \label{fig:neighbours_scores}
\end{figure}

Our derivation of the message-passing algorithm given in Eq.~(\ref{eq_truncated_posterior}) started by the replacement of the posterior distribution by its truncated version; once this approximation had been made we wrote the truncated posterior in terms of the probability laws $\mathbb{P}_1^{(d)}$ and $\mathbb{P}_0^{(d)}$, which is asymptotically exact according to the local analysis of the correlated random graph ensemble. Finally, we constructed an estimator by maximizing the marginal posterior probability. If the full posterior distribution were used, this procedure would maximize the average overlap with the ground truth. The scores derived in this way can be interpreted as the likelihood ratios of an hypothesis testing problem between correlated and uncorrelated pairs of trees, even if this was not used in the derivation of the algorithm itself. This perspective motivates further investigations on the properties of the neighborhoods compared by the algorithm. For example, some pairs of neighborhoods in the graphs are neither drawn from the correlated law $\mathbb{P}_1^{(d)}$ nor from the uncorrelated product of the laws $\mathbb{P}_0^{(d)}$. To be more precise, let us consider a vertex $i$ of the graph $\bA$, its image $\bpis(i)$ in $\bB$ under the groundtruth permutation, and a neighbor of the latter $i' \in \partial \bpis(i)$; we shall call in the following $(i,i')$ a \textit{quasi-aligned pair of vertices}. It should be clear that the law of the neighborhoods $(T_i,T'_{i'})$ is neither $\mathbb{P}_1^{(d)}$, because $i' \neq \bpis(i)$, nor the product of $\mathbb{P}_0^{(d)}$, because $i$ and $i'$ correspond to vertices at distance 1 in the colored graph $\bG$, hence their neighborhoods overlap in general via the bicolored edges of $\bG$. The presence of this correlation could suggest that the scores $L_{i,\bpis(i)}$ of an aligned pair and $L_{i,i'}$ with $i' \in \partial \bpis(i)$ of a quasi-aligned pair are of the same order, hence that the algorithm could easily mistake one for the other. In order to investigate the possible presence of an issue due to this correlation, we estimated the typical values of the scores for aligned and quasi-aligned pairs of vertices, by computing the following averages of the logarithm of the scores:
\begin{align}
    \label{eq:argmax_score}
    & A_1 = \mathbb{E}\left[ \frac{1}{n}\sum_{i=1}^n \ln L^{(d)}_{i,\bpih(i)}\right] \ , \\
    \label{eq: ground_truth_score}
    & A_2 = \mathbb{E}\left[\frac{1}{n}\sum_{i=1}^n \ln L^{(d)}_{i,\bpis(i)}\right] \ , \\
    \label{eq:neigh_B_score}
    & A_3 = \mathbb{E}\left[\frac{1}{n}\sum_{i=1}^n\frac{1}{|\partial \bpis(i)|}\sum_{i'\in\partial \bpis(i)}\ln L^{(d)}_{i,i'}\right] \ , \\
    \label{eq:rand_score}
    & A_4 = \mathbb{E}\left[\frac{1}{n^2}\sum_{i,i' =1}^n \ln L^{(d)}_{i,i'}\right]  \ .
\end{align}
The averages are intended over the graphs realizations and over pairs of vertices which are, respectively, the ones selected by the estimator, $A_1$, the ones in the groundtruth, $A_2$, the quasi-aligned ones $A_3$, and finally randomly selected pairs $A_4$. The numerical estimation of these averages is presented in Fig.~\ref{fig:neighbours_scores}, which shows that $A_1 \ge A_2 \gg A_3 \approx A_4$. The fact that $A_2 \gg A_3 \approx A_4$ is very reassuring and should dissipate the concern raised above: pairs of quasi-aligned vertices have scores substantially smaller than the aligned ones, which avoids the possible confusion between them, and of the same order as arbitrarily distant vertices. An interpretation of this result proceeds as follows. The function $L(T,T')$ is, forgetting its precise definition, a function of the structure of the rooted trees $T$ and $T'$. Given a tree $T$ and two adjacent vertices $a$ and $b$, the ordered structure of the same tree $T$ rooted in $a$ is very different from the one of the tree rooted in $b$ (the roots themselves might have different degrees, and no common subtree might appear), hence a priori the two structures have very different images under the function $L(\cdot,T')$. Interestingly, this argument breaks down if $a$ and $b$ are vertices at distance 2: consider indeed the path $a-c-b$ in $T$, the tree rooted at $a$ and the tree rooted at $b$. The two trees will share a subtree rooted in $c$. We have checked that indeed the average of the logarithm of the scores between $i$ and a vertex at distance $2$ from $\bpis(i)$ in $B$ is larger than $A_3$, but still notably smaller than $A_2$.

\subsection{Alternative estimators}
\label{app:matrix_estimator}

As we discussed at the beginning of Section~\ref{sec:inference}, the notion of optimal estimator in inference problems depends on the choice of the distance between the estimator and the signal that is to be minimized. In the main text we chose an estimator $\pih$ that, for each node $i$ of the graph $A$, maximizes the (approximate) probability of the event $\pih(i)=\bpis(i)$, aiming at maximizing the average overlap between $\bpih$ and $\bpis$, as defined in \eqref{eq_overlap}. We will consider here other estimators, devised to optimize other error measures. 

Before giving some examples, let us introduce additional definitions that will be useful in the discussion. We shall consider partial estimators, namely functions $\pih$ from $[n]$ to $[n] \cup \{ * \}$, where $*$ is an additional dummy symbol, such that $\pih(i)=*$ whenever the partial estimator does not propose any vertex of the graph $B$ to be matched with the vertex $i$ of $A$. We shall denote $\mathcal{S}(\pih) = \{ i \in [n] \ : \ \pih(i) \neq * \}$ the set of vertices that are assigned by $\pih$, $n(\pih) = |\mathcal{S}(\pih)|$ their number, and define the overlap of a partial estimator $\pih$ with a permutation $\pi_*$ of $[n]$ as
\begin{equation}
    \ov(\pih,\pi_*) = \frac{1}{n(\pih)} \sum_{i \in \mathcal{S}(\pih)} \mathbb{I}( \pih(i) = \pi_*(i) ) \ .
\label{eq_overlap_partial}
\end{equation}
This counts the fraction of correct matches among the assigned ones, and coincides with the definition in Eq.~(\ref{eq_overlap}) when $\pih$ assigns all the vertices.

Let us also recall the notation $P_{i,i'}=\mathbb P(\bpis(i) = i' |\bA=A,\bB=B)$ introduced in Section~\ref{sec:inference}, whose row and column sums are normalized to 1, and in which we keep implicit the dependency on the observed graphs $A$ and $B$. As this marginal of the exact posterior is not efficiently computable we will use its approximation $\widehat{P}_{i,i'}$ obtained from the message passing algorithm. According to its derivation in terms of the truncated posterior, $\widehat{P}_{i,i'}$ should be proportional to the score matrix $L^{(d)}_{i,i'}$; to ensure one type of normalization we will define
\begin{equation}
 \label{eq:mat_approx_phat}
   \widehat{P}_{i,i'}  = \frac{L^{(d)}_{i,i'}}{\sum_{j'} L^{(d)}_{i,j'}} \ ,
\end{equation}
which by definition satisfies the same row normalization as the exact quantity $P$, but because of the approximation may violate the column normalization.

\paragraph{Matrix estimator}

Let us consider the following distance (or loss function) between a (possibly partial) estimator $\pih$ and a permutation $\pi_*$:
\begin{equation}
    \mathcal{L}(\pih,\pi_*)=\left(1-\frac{n(\pih)}{n}\right)+\frac{2 n(\pih)}{n}\left(1-\ov(\pih,\pi_*)\right) \ ,
\label{eq_loss_L}
\end{equation}
which coincides with the Hamming distance between the matrix representation of $\pih$ and $\pi_*$ as discussed in Eq.~(\ref{eq_matrix_estimator}) (the Hamming distance between an empty row and a row containing exactly one 1 is 1, while it is 2 between two rows containing exactly one 1 at different positions). Minimizing $\mathcal{L}$ amounts to find a compromise between the two terms: the first one favors estimators that assign the largest possible number of vertices, but the second one grows if too many of these assignment are erroneous. The error $\mathcal{L}$ satisfies the bounds $0\leq \mathcal{L}(\pih,\pi_*)\leq 2$, with $\mathcal{L}=2$ if and only if  all the vertices are assigned and are all incorrect, and $\mathcal{L}=1$ for the null estimator which does not assign any of the vertices.

\begin{figure}
    \centering
    \includegraphics[width=\columnwidth]{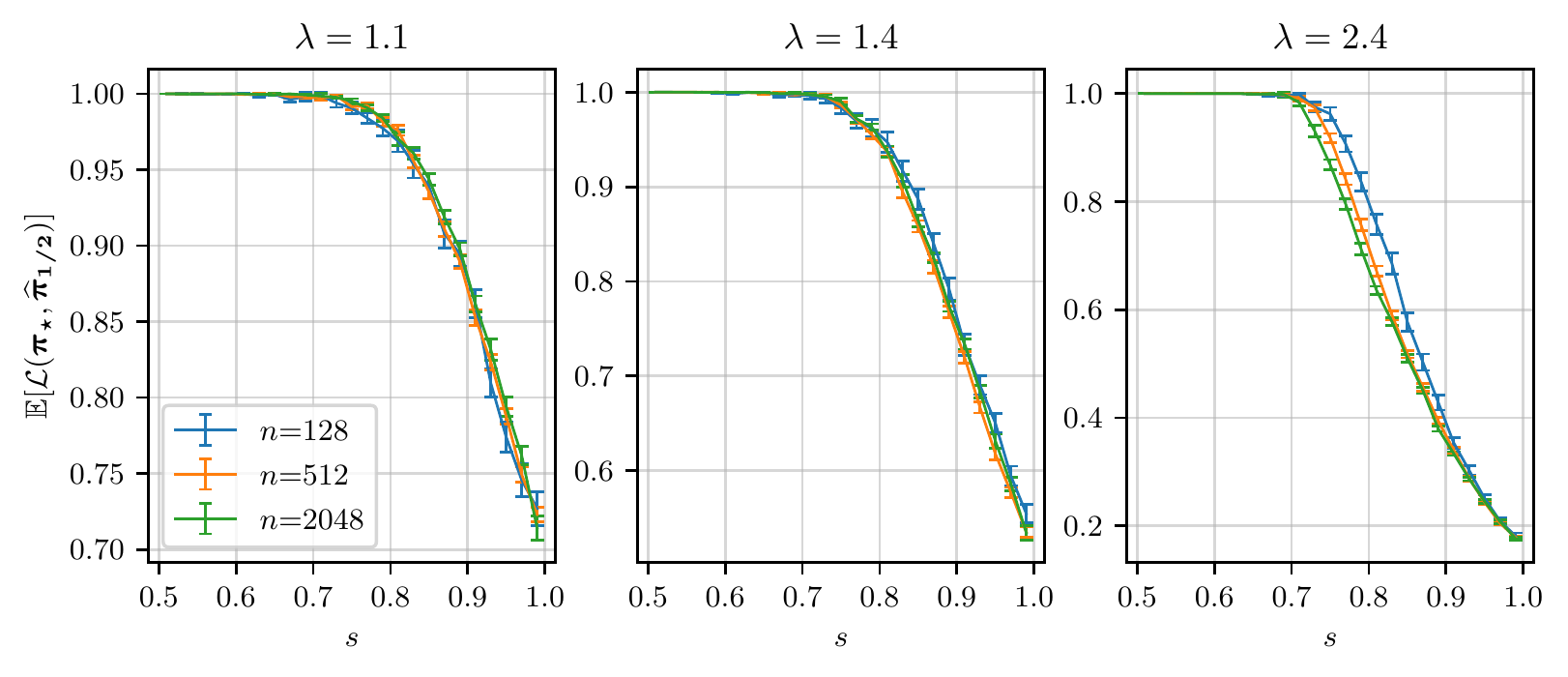}
    \caption{The average loss $\mathbb{E}[\mathcal L(\bpih_{1/2},\bpis)]$ as a function of the correlation $s$ for different values of the system size $n$. In increasing order of $n$ we averaged over $100,100,25$ independent realizations of the two graphs $\bA,\bB$.}
    \label{fig:matrix_estimator_error_scalingN}
\end{figure}

The estimator that minimizes this loss on average is built from the posterior probabilities $P_{i,i'}$ as
\begin{equation}
    \bpih_{1/2}(i) = \begin{cases} i' & \text{if} \quad P_{i,i'} > \frac{1}{2} \\
    * & \text{if} \quad P_{i,j'} \le \frac{1}{2} \ \ \forall \ j'
    \end{cases} \ ;
\label{eq_pih_12}
\end{equation}
because of the normalization condition at most one $i'$ can be selected in the first line. The numerical results presented in Fig.~\ref{fig:matrix_estimator_error_scalingN} have been obtained by replacing in this expression $P$ by its approximation $\widehat{P}$ defined in Eq.~(\ref{eq:mat_approx_phat}), and selecting for each set of parameters the optimal value of the depth $d$ (the one that minimized the average loss). The curves in Fig.~\ref{fig:matrix_estimator_error_scalingN} shows the average error $\mathcal{L}$ between this estimator and the groundtruth as a function of $s$, for different values of $n$. The error decreases below $\mathcal{L}=1$ for values of $s$ around $0.7$, slightly above the threshold $s_{\rm algo} \approx 0.6$ we observed in the main part of the text for the estimator $\bpih$: as a matter of fact in between these two values of $s$ the algorithm is not confident enough about the quality of its predictions (in more technical terms all $\widehat{P}_{i,i'}$ are below $1/2$), hence it prefers to return the null estimator instead of some possibly erroneous matches. One can also observe that the finite size effects seem to be much weaker for this estimator than for $\bpih$, as the curves for various $n$ are almost superimposed in Fig.~\ref{fig:matrix_estimator_error_scalingN}, to be compared for instance with Fig.~\ref{fig:optimal_ovlap}.

\paragraph{Thresholded estimator}

We will now discuss a generalization of the estimator $\boldsymbol \pih_{1/2}$ that amounts in some sense to tune the relative weights of the two terms in the loss function (\ref{eq_loss_L}). Indeed in some applications it might be preferable to propose a large number of matched vertices, at the risk of making many mistakes, or on the contrary to return a very partial estimator but with a large overlap for the few assigned vertices. This point was formalized under the name of one-sided partial recovery in~\cite{GaMaLe21b}; we recall that in the sparse regime we are investigating some errors are unavoidable, in particular because of an extensive number of isolated vertices in the bicolored graph.

For a given ``budget'' of vertices to assign one should make a choice of which vertices to put in the estimator, and it would be advisable to select the ones which are the most likely to be correctly matched. As $\max_{i'} P_{i,i'}$ is the probability (over $\bpis$, conditional on $\bA,\bB$) that $\bpih(i)=\bpis(i)$, the information theoretically optimal choice (if one has access to the exact posterior distribution) is thus to consider a threshold $T \in [0,1]$ and define the following estimator:
\begin{equation}
    \boldsymbol \pih_{T}(i) = \begin{cases} \arg\max_{i'} P_{i,i'} & \text{if} \quad \max_{i'} P_{i,i'} > T \\
    * & \text{if} \quad \max_{i'} P_{i,i'} \le T 
    \end{cases} \ ,
\label{eq_estimator_threshold}
\end{equation}
that coincides with (\ref{eq_pih_12}) when $T=1/2$, thus justifying the notation. When $T=0$ one recovers from this formula the full estimator $\pih$ of the main text, and for general values of $T$ one has $\boldsymbol \pih_T(i)=\bpih(i)$ for all the assigned vertices. The role of $T$ is thus to control the number of assigned vertices in the partial estimator, larger values of $T$ corresponding to smaller values of $n(\pih_T)$, the estimator concentrating on the vertices which are the most likely to be correctly matched.

\begin{figure}
    \centering
    \includegraphics[width=\columnwidth]{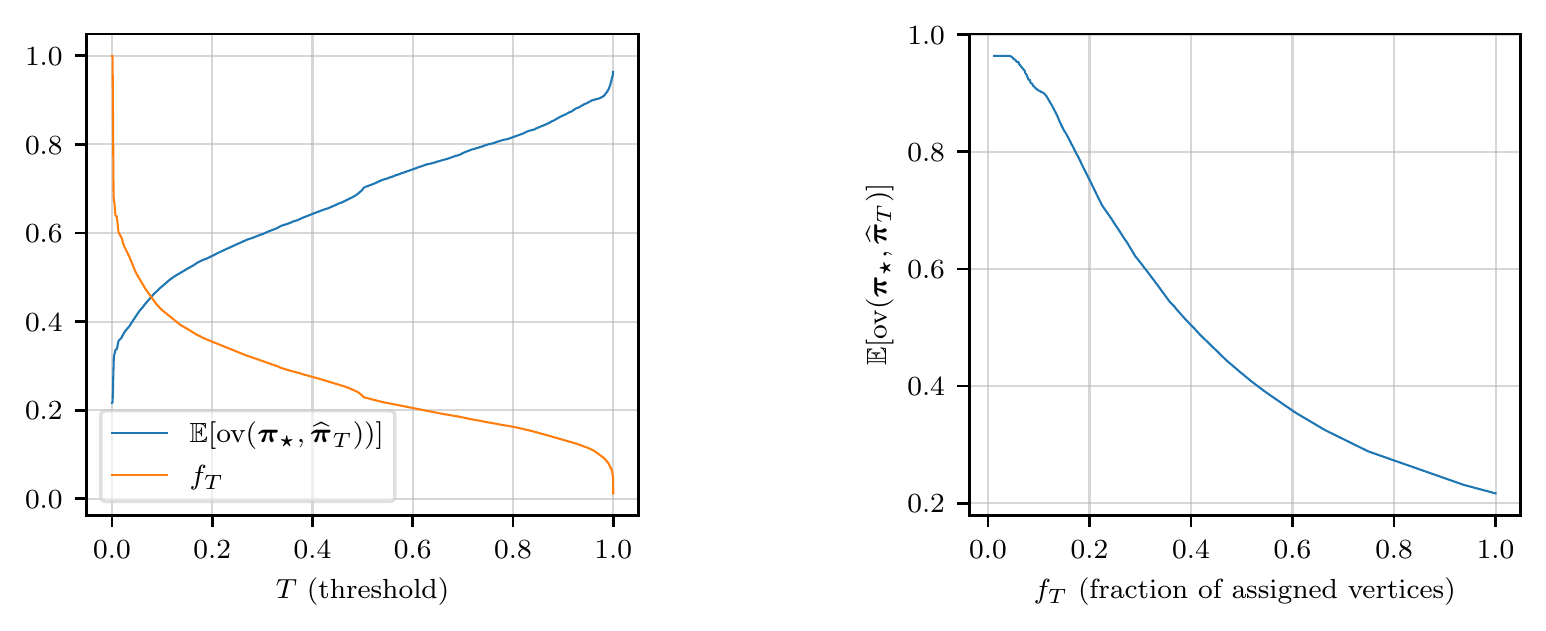}
    \caption{Performance of the threshold estimator of Eq.~\eqref{eq_estimator_threshold} for system size $n=2048$, average degree $\lambda=1.4$, correlation $s=0.83$, depth $d=10$, averages being taken over $25$ independent realizations of $(\bA, \bB)$. (Left) Average overlap and fraction of assigned vertices $f_T$ as a function of the threshold: with larger $T$ fewer vertices are included in the partial alignment but the overlap increases.  (Right) Average overlap as function of the fraction of assigned vertices. The plot is obtained from the left panel: increasing $T$ from zero to one corresponds to following the curve from the bottom right corner to the top left one. 
    }
    \label{fig:threshold_estimator_v2}
\end{figure}

In practice we used this formula replacing $P$ by its proxy $\widehat{P}$ computed from the message-passing algorithm according to Eq.~\eqref{eq:mat_approx_phat}, and obtained in this way the curves of Fig.~\ref{fig:threshold_estimator_v2}. On the left panel we present the fraction of assigned vertices $f_T=\mathbb{E}[n(\bpih_T)]/n$ and the corresponding overlap (computed only among the assigned vertices according to Eq.~\eqref{eq_overlap_partial}) as a function of the threshold $T$. For small $T$ most vertices are included in the partial alignment, however the overlap is small and close to that of the full estimator $\bpih$. When $T$ increases the fraction of assigned vertices is reduced, but the partial overlap increases, showing that despite the approximation incurred when replacing $P$ by $\widehat{P}$ the algorithm is indeed able to select the vertices that it manages to align correctly. In particular for $T\to1$ only a few vertices are included, but they are almost all matched correctly (the overlap is close to $1$). The left panel is a parametric representation of the same data, displaying the overlap as a function of the fraction of assigned vertices, 
hence the trade-off between the two opposite requirements of predicting a match for a large number of vertices, and doing this accurately.

\subsection{A consistency check of the approximation}
\label{app:nishimori}

We will now present some numerical tests of the accuracy of the approximation made by replacing the posterior probabilities $P_{i,i'}$ by the expression $\widehat{P}_{i,i'}$ of Eq.~\eqref{eq:mat_approx_phat}. Both quantities depend implicitly on the observed pair of graphs $A,B$. Let us start considering the optimal estimator $\bpih_{\rm opt}$ defined from the exact posterior probabilities as $\bpih_{\rm opt}(i)=\arg\max_{i'} P_{i,i'}$. Its average overlap with the ground-truth can be written in two equivalent ways,
\begin{equation}\begin{split}
    \mathbb{E}[\ov(\bpih_{\rm opt},\bpis)] & = \frac{1}{n} \sum_{i=1}^n \mathbb{E}[ \mathbb{I}(\bpis(i) = \arg\max_{i'} P_{i,i'} )  ] = \frac{1}{n} \sum_{i=1}^n \mathbb{E}\left[ \sum_{j'} P_{i,j'} \mathbb{I}(j' = \arg\max_{i'} P_{i,i'} ) \right]  \\ & = \frac{1}{n} \sum_{i=1}^n \mathbb{E}[\max_{i'} P_{i,i'} ] \ ,
\end{split}\end{equation}
where we used the fact, called \textit{Nishimori condition} \cite{zdeborova2016statistical}, that the law of $(\bpis,(\bA,\bB))$ is the same as the law of $(\bpi',(\bA,\bB))$ if $\bpi'$ is drawn from the posterior law given $(\bA,\bB)$. Note that in the last expression the ground-truth expression does not appear explicitly.

We turn now to the estimator $\bpih(i)=\arg\max_{i'} \widehat{P}_{i,i'}$ based on the approximation of the posterior probabilities through the message-passing algorithm, which is the only one we can use in practice, and define what we shall call the true average overlap,
\begin{equation}
    \ov = \mathbb{E}[\ov(\bpih,\bpis)] = \frac{1}{n} \sum_{i=1}^n \mathbb{E}[ \mathbb{I}(\bpis(i) = \arg\max_{i'} \widehat{P}_{i,i'} )  ] \ ,
\label{eq_ov}
\end{equation}
and the estimated average overlap,
\begin{equation}
    \widehat{\ov} = \frac{1}{n} \sum_{i=1}^n \mathbb{E}[\max_{i'} \widehat{P}_{i,i'} ] \ ,
\label{eq_hov}
\end{equation}
where in the last equation we do not use explicitly the knowledge of the ground-truth permutation, $\max_{i'} \widehat{P}_{i,i'}$ being the estimation by the algorithm itself of the probability that its prediction $\bpih(i)$ is correct. According to the discussion above we would have $\ov=\widehat{\ov}$ if the approximated probabilities $\widehat{P}_{i,i'}$ coincided with the exact ones $P_{i,i'}$.

\begin{figure}
    \centering
    \includegraphics{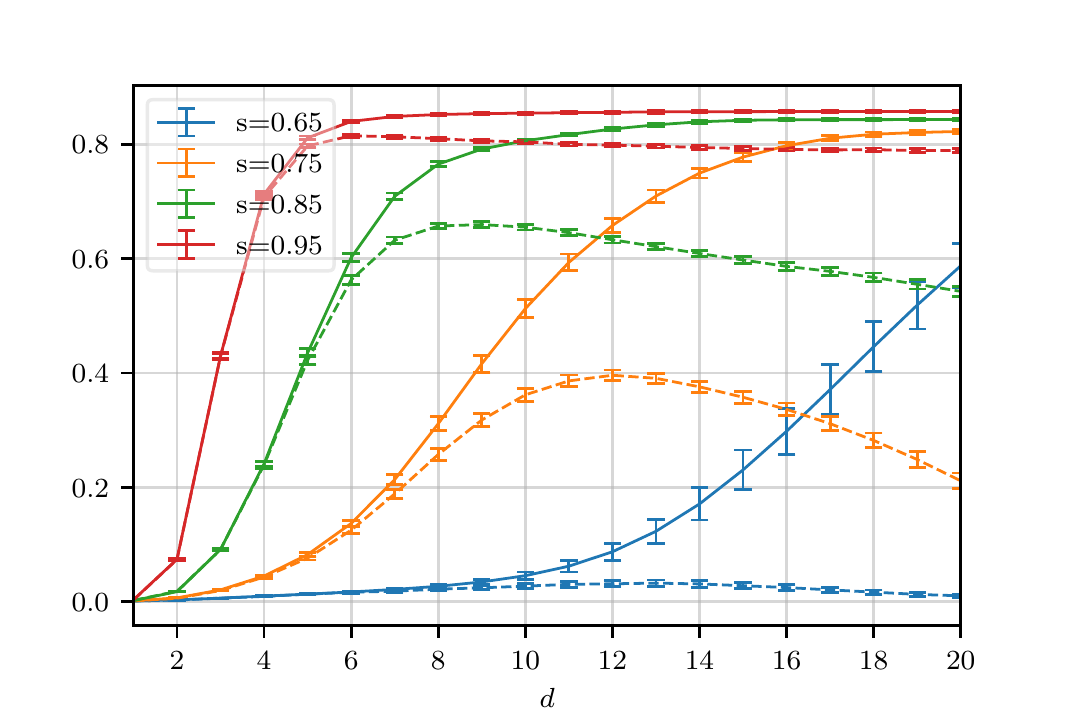}
    \includegraphics[width=\columnwidth]{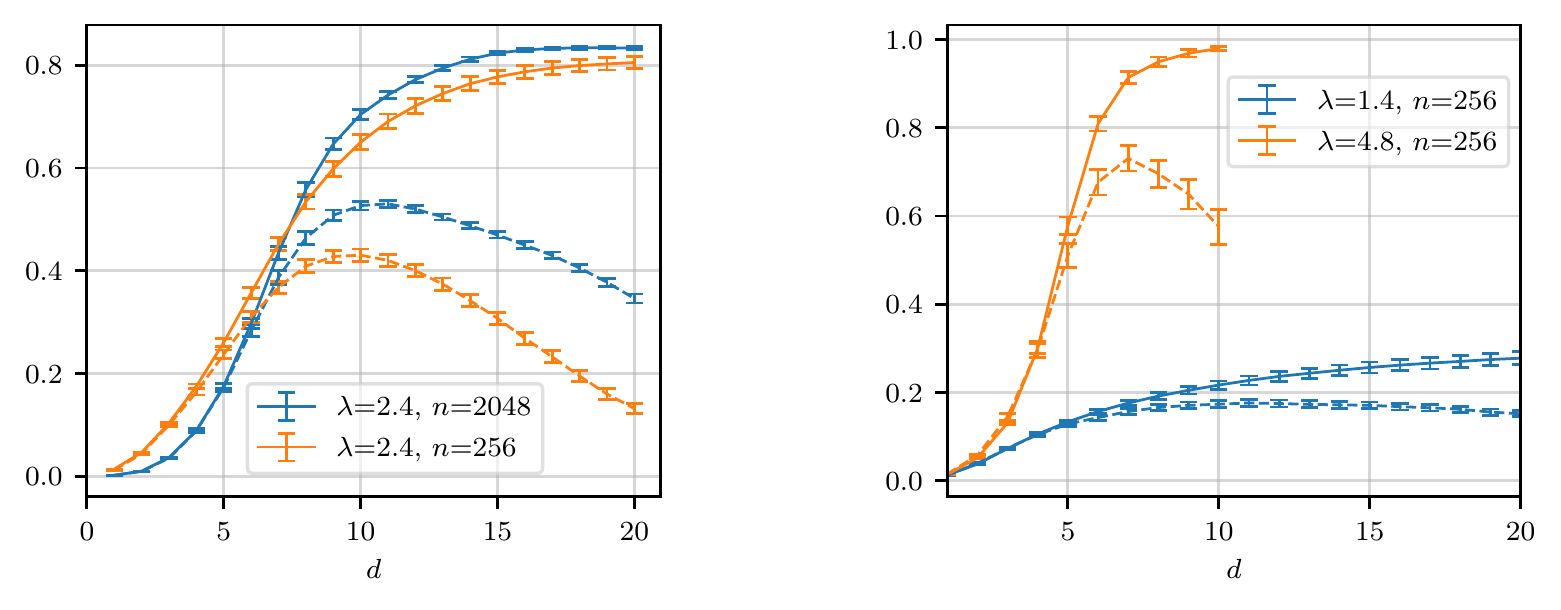}
    \caption{A comparison of the true average overlap (dashed line) of Eq.~(\ref{eq_ov}) and of the estimated overlap (solid line) of Eq.~(\ref{eq_hov}).
    (Top) $n=2048$, $\lambda=2.4$: for small $d$ the estimated and true overlap agree, however increasing $d$ the estimated overlap grows towards $1$ showing that the algorithm is overconfident of its predictions. (Bottom left) when $n$ is increased the value of $d$ at which the true and estimated overlap ceases to agree also increases. (Bottom right) when $\lambda$ is increased the value of $d$ at which the 
    two curves separates decreases, possibly because of stronger finite size effects for larger $\lambda$.}
    \label{fig:estimated_vs_true_ovlap_n2048}
\end{figure}

We present in Fig.~\ref{fig:estimated_vs_true_ovlap_n2048} the numerical results of the comparison of the two quantities $\ov$ and $\widehat{\ov}$; to account for the dependency of $\widehat{P}$ on the depth parameter $d$ of the message-passing algorithm we plot both quantities as a function of $d$ for fixed values of $n,\lambda,s$. One can see on the figure that they coincide for small $d$ but differ from each other for larger values of $d$; as a matter of fact when $d$ increases the estimated probability laws $\widehat{P}_{i,\cdot}$ become more and more concentrated on their mode $\bpih(i)$, i.e., close to $\widehat{P}_{i,i'}=\delta_{i',\bpih(i)}$. This means that the algorithm becomes overconfident about its predictions, its estimation of the probability that the predicted match is correct is larger than the true one. The plots on the bottom of Fig.~\ref{fig:estimated_vs_true_ovlap_n2048} study the dependency of the separation point between the two curves on $\lambda$ and $n$. Larger $n$ leads to agreement up to higher values of $d$, while higher $\lambda$ makes the two curves depart at smaller $d$. Let us finally underline that if $P=\hat P$ implies that $\ov=\widehat{\ov}$, the converse implication is far from being true: suppose for instance that $\widehat{P}_{i,i'}=1/n$ for all $i,i'$, i.e., that the estimation does not extract any information from the observations and that $\bpih(i)$ is a uniformly random vertex (the ties in the $\arg\max$ being broken at random). Then from \eqref{eq_ov} and \eqref{eq_hov} it follows that $\ov=\widehat{\ov}=1/n$. However in a hard phase where partial recovery is information theoretically possible this situation would occur with a non-trivial $P \neq \widehat{P}$.

\section*{Bibliography}
\bibliography{biblio.bib}

\providecommand{\newblock}{}
\begin{thebibliography}{10}
\expandafter\ifx\csname url\endcsname\relax
  \def\url#1{{\tt #1}}\fi
\expandafter\ifx\csname urlprefix\endcsname\relax\def\urlprefix{URL }\fi
\providecommand{\eprint}[2][arXiv]{#1:\linebreak[0]#2}

\bibitem{Conte2004}
Conte D, Foggia P, Sansone C and Vento M 2004 Thirty years of graph matching in
  pattern recognition {\em International journal of pattern recognition and
  artificial intelligence\/} {\bf 18} 265--298

\bibitem{Narayanan2008}
Narayanan A and Shmatikov V 2008 Robust de-anonymization of large sparse
  datasets {\em 2008 IEEE Symposium on Security and Privacy (sp 2008)\/} (IEEE)
  pp 111--125

\bibitem{Pedarsani2011}
Pedarsani P and Grossglauser M 2011 On the privacy of anonymized networks {\em
  Proceedings of the 17th ACM SIGKDD International Conference on Knowledge
  Discovery and Data Mining\/} KDD '11 (New York, NY, USA: Association for
  Computing Machinery) p 1235–1243 ISBN 9781450308137

\bibitem{ramani2003exploiting}
Ramani A~K and Marcotte E~M 2003 Exploiting the co-evolution of interacting
  proteins to discover interaction specificity {\em Journal of molecular
  biology\/} {\bf 327} 273--284

\bibitem{berg2006cross}
Berg J and L{\"a}ssig M 2006 {Cross-species analysis of biological networks by
  Bayesian alignment} {\em Proceedings of the National Academy of Sciences\/}
  {\bf 103} 10967--10972

\bibitem{li2007alignment}
Li Z, Zhang S, Wang Y, Zhang X~S and Chen L 2007 Alignment of molecular
  networks by integer quadratic programming {\em Bioinformatics\/} {\bf 23}
  1631--1639

\bibitem{singh2008global}
Singh R, Xu J and Berger B 2008 Global alignment of multiple protein
  interaction networks with application to functional orthology detection {\em
  Proceedings of the National Academy of Sciences\/} {\bf 105} 12763--12768

\bibitem{nowak2018revised}
Nowak A, Villar S, Bandeira A~S and Bruna J 2018 Revised note on learning
  quadratic assignment with graph neural networks {\em 2018 IEEE Data Science
  Workshop (DSW)\/} (IEEE) pp 1--5

\bibitem{azizian2020expressive}
Azizian W and Lelarge M 2020 Expressive power of invariant and equivariant
  graph neural networks {\em arXiv:2006.15646\/}

\bibitem{Burkard1998}
Burkard R~E, {\c{C}}ela E, Pardalos P~M and Pitsoulis L~S 1998 {\em The
  Quadratic Assignment Problem\/} (Boston, MA: Springer US) pp 1713--1809 ISBN
  978-1-4613-0303-9

\bibitem{Cullina2016}
Cullina D and Kiyavash N 2016 Improved achievability and converse bounds for
  erdos-renyi graph matching {\em Proceedings of the 2016 ACM SIGMETRICS
  International Conference on Measurement and Modeling of Computer Science\/}
  SIGMETRICS `16 (New York, NY, USA: Association for Computing Machinery) p
  63–72 ISBN 9781450342667

\bibitem{Ding2018}
Ding J, Ma Z, Wu Y and Xu J 2021 Efficient random graph matching via degree
  profiles {\em Probability Theory and Related Fields\/} {\bf 179}(1) 29--115

\bibitem{Fan2019}
Fan Z, Mao C, Wu Y and Xu J 2020 Spectral graph matching and regularized
  quadratic relaxations: Algorithm and theory {\em Proceedings of the 37th
  International Conference on Machine Learning\/} ({\em Proceedings of Machine
  Learning Research\/} vol 119) ed III H~D and Singh A (PMLR) pp 2985--2995

\bibitem{MaRuTi21}
Mao C, Rudelson M and Tikhomirov K 2021 Exact matching of random graphs with
  constant correlation {\em arXiv:2110.05000\/}

\bibitem{GaMa20}
Ganassali L and Massouli\'e L 2020 From tree matching to sparse graph alignment
  {\em Proceedings of Thirty Third Conference on Learning Theory\/} ({\em
  Proceedings of Machine Learning Research\/} vol 125) (PMLR) pp 1633--1665

\bibitem{WuXuYu21}
Wu Y, Xu J and Yu S~H 2021 Settling the sharp reconstruction thresholds of
  random graph matching {\em arXiv:2102.00082\/}

\bibitem{GaMaLe21a}
Ganassali L, Massouli\'e L and Lelarge M 2021 Impossibility of partial recovery
  in the graph alignment problem {\em Proceedings of Thirty Fourth Conference
  on Learning Theory\/} ({\em Proceedings of Machine Learning Research\/} vol
  134) (PMLR) pp 2080--2102

\bibitem{GaMaLe21b}
Ganassali L, Massouli\'e L and Lelarge M 2021 Correlation detection in trees
  for partial graph alignment {\em arXiv:2107.07623\/}

\bibitem{Hall2020}
Hall G and Massouli\'e L 2020 {Partial Recovery in the Graph Alignment Problem}
  {\em arxiv:2007.00533\/}

\bibitem{OnGaEr16}
Onaran E, Garg S and Erkip E 2016 Optimal de-anonymization in random graphs
  with community structure {\em 2016 50th Asilomar Conference on Signals,
  Systems and Computers\/} pp 709--713

\bibitem{RaSr21}
Racz M~Z and Sridhar A 2021 Correlated stochastic block models: Exact graph
  matching with applications to recovering communities {\em arxiv:2107.06767\/}

\bibitem{GaLeMa19}
Ganassali L, Lelarge M and Massouli\'e L 2019 {Spectral Alignment of Correlated
  Gaussian matrices} {\em arXiv:1912.00231\/}

\bibitem{Ga20}
Ganassali L 2020 {Sharp threshold for alignment of graph databases with
  Gaussian weights} {\em arXiv:2010.16295\/}

\bibitem{BrBrMaTrWeZe10}
Bradde S, Braunstein A, Mahmoudi H, Tria F, Weigt M and Zecchina R 2010
  Aligning graphs and finding substructures by a cavity approach {\em {EPL}
  (Europhysics Letters)\/} {\bf 89} 37009

\bibitem{BaGlSaWa13}
Bayati M, Gleich D~F, Saberi A and Wang Y 2013 Message-passing algorithms for
  sparse network alignment {\em ACM Trans. Knowl. Discov. Data\/} {\bf 7}

\bibitem{Al01}
Aldous D~J 2001 The $\zeta(2)$ limit in the random assignment problem {\em
  Random Structures \& Algorithms\/} {\bf 18} 381--418

\bibitem{MePa01}
M{\'e}zard M and Parisi G 2001 The bethe lattice spin glass revisited {\em Eur.
  Phys. J. B\/} {\bf 20} 217

\bibitem{DeKrMoZd11}
Decelle A, Krzakala F, Moore C and Zdeborov\'a L 2011 Asymptotic analysis of
  the stochastic block model for modular networks and its algorithmic
  applications {\em Phys. Rev. E\/} {\bf 84}(6) 066106

\bibitem{Mo17}
Moore C 2017 The computer science and physics of community detection:
  Landscapes, phase transitions, and hardness {\em Bulletin of EATCS\/} {\bf 1}
  preprint: arXiv:1702.00467

\bibitem{Ab18}
Abbe E 2018 Community detection and stochastic block models: Recent
  developments {\em Journal of Machine Learning Research\/} {\bf 18} 1--86

\bibitem{MoNeSl16}
Mossel E, Neeman J and Sly A 2016 Belief propagation, robust reconstruction and
  optimal recovery of block models {\em Ann. Appl. Probab.\/} {\bf 26}
  2211--2256

\bibitem{MaWuXuYu21}
Mao C, Wu Y, Xu J and Yu S~H 2021 Testing network correlation efficiently via
  counting trees {\em arxiv:2110.11816\/}

\bibitem{Ot48}
Otter R 1948 The number of trees {\em Annals of Mathematics\/} {\bf 49}
  583--599

\bibitem{KiSuVu02}
Kim J~H, Sudakov B and Vu V~H 2002 On the asymmetry of random regular graphs
  and random graphs {\em Random Structures \& Algorithms\/} {\bf 21} 216--224

\bibitem{YaGr13}
Yartseva L and Grossglauser M 2013 On the performance of percolation graph
  matching {\em Proceedings of the First ACM Conference on Online Social
  Networks\/} COSN '13 (New York, NY, USA: Association for Computing Machinery)
  p 119–130

\bibitem{LyFiPr14}
Lyzinski V, Fishkind D~E and Priebe C~E 2014 Seeded graph matching for
  correlated erd\"{o}s-r\'{e}nyi graphs {\em J. Mach. Learn. Res.\/} {\bf 15}
  3513–3540

\bibitem{MoXu20}
Mossel E and Xu J 2020 Seeded graph matching via large neighborhood statistics
  {\em Random Structures \& Algorithms\/} {\bf 57} 570--611

\bibitem{YuXuLi21}
Yu L, Xu J and Lin X 2021 The power of $d$-hops in matching power-law graphs
  {\em arxiv:2102.12975\/}

\bibitem{zdeborova2016statistical}
Zdeborov{\'a} L and Krzakala F 2016 Statistical physics of inference:
  Thresholds and algorithms {\em Advances in Physics\/} {\bf 65} 453--552

\end{thebibliography}
\end{document}